\documentclass[sts]{imsart}

\RequirePackage[OT1]{fontenc}
\usepackage{amsthm,amsmath,natbib}
\RequirePackage[colorlinks,citecolor=blue,urlcolor=blue]{hyperref}

\usepackage[dvipsnames,table]{xcolor}
\usepackage{graphicx}
\usepackage{verbatim}
\usepackage{array}
\usepackage{booktabs} 
\usepackage{longtable}

\usepackage[font=small,labelfont=bf,singlelinecheck=false]{caption} 
\usepackage{wrapfig} 

\startlocaldefs
\numberwithin{equation}{section}
\theoremstyle{plain}

\endlocaldefs

\newcommand{\E}{\mathrm{E}}

\newcommand{\knob}{$\left[\text{knob}\right]$ }
\newcommand{\oracle}{$\left[\text{oracle}\right]$ }

\makeatletter
\def\note#1{\def\tempa{#1}\futurelet\next\note@i}
\def\note@i{\ifx\next\bgroup\expandafter\note@ii\else\expandafter\note@end\fi}
\def\note@ii#1{\textcolor{red}{\tempa\ : #1}}
\def\note@end{\textcolor{red}{: \tempa}}
\makeatother

\makeatletter
\def\todo#1{\def\tempa{#1}\futurelet\next\todo@i}
\def\todo@i{\ifx\next\bgroup\expandafter\todo@ii\else\expandafter\todo@end\fi}
\def\todo@ii#1{\textcolor{red}{\tempa\ TODO: #1}}
\def\todo@end{\textcolor{red}{TODO: \tempa}}
\makeatother

\definecolor{lightgray}{gray}{0.975}

\begin{document}

\begin{frontmatter}
\title{Automated versus do-it-yourself methods for causal inference: Lessons learned from a data analysis competition\thanksref{T1}}
\runtitle{Causal Inference Competition}
\thankstext{T1}{This research was partially supported by Institute of Education Sciences grants R305D110037 and R305B120017}

\begin{aug}
\author{\fnms{Vincent} \snm{Dorie}\thanksref{t1}\ead[label=e1]{vdorie@gmail.com}},
\author{\fnms{Jennifer} \snm{Hill}\ead[label=e2]{jennifer.hill@nyu}},
\author{\fnms{Uri} \snm{Shalit}\ead[label=e3]{urish22@gmail.com}},
\author{\fnms{Marc} \snm{Scott}\ead[label=e4]{marc.scott@nyu.edu}},
\and
\author{\fnms{Dan} \snm{Cervone}\ead[label=e5]{dcervone@gmail.com}
\ead[label=u1,url]{www.foo.com}}

\thankstext{t1}{Vincent Dorie is a postdoctoral researcher (Q-Train) at New York University, Jennifer Hill is Professor of Applied Statistics and Data Science at New York University, Uri Shalit is an Assistant Professor at Technion, Marc Scott is Professor of Applied Statistics at New York University, Dan Cervone is Senior Analyst in Research and Development with the Los Angeles Dodgers}

\runauthor{V. Dorie et al.}

\affiliation{New York University}
\end{aug}

\begin{abstract}
Statisticians have made great progress in creating methods that reduce our reliance on parametric assumptions. However this explosion in research has resulted in a breadth of inferential strategies that both create opportunities for more reliable inference as well as complicate the choices that an applied researcher has to make and defend. Relatedly, researchers advocating for new methods typically compare their method to at best 2 or 3 other causal inference strategies and test using simulations that may or may not be designed to equally tease out flaws in all the competing methods. The causal inference data analysis challenge, ``Is Your SATT Where It's At?'', launched as part of the 2016 Atlantic Causal Inference Conference, sought to make progress with respect to both of these issues. The researchers creating the data testing grounds were distinct from the researchers submitting methods whose efficacy would be evaluated. Results from 30 competitors across the two versions of the competition (black box algorithms and do-it-yourself analyses) are presented along with post-hoc analyses that reveal information about the characteristics of causal inference strategies and settings that affect performance. The most consistent conclusion was that methods that flexibly model the response surface perform better overall than methods that fail to do so.  Finally new methods are proposed that combine features of several of the top-performing submitted methods.
\end{abstract}


\end{frontmatter}

\section{Introduction}

In the absence of a controlled randomized or natural experiment,\footnote{We use natural experiment to include 1) studies where the causal variable is randomized not for the purposes of a study (for instance a school lottery), 2) studies where a variable is randomized but the causal variable of interest is downstream of this (e.g. plays the role of an instrumental variable), and 3) regression discontinuity designs.} inferring causal effects involves the difficult task of constructing fair comparisons between observations in the control and treatment groups.  Since these groups can differ in substantive and non-obvious ways  researchers are incentivized to control for a large number of pre-treatment covariates.\footnote{We note that some have cautioned against this temptation due to the potential for some variables to amplify the bias that remains when ignorability is not satisfied \citep{pear:2010,midd:etal:2016,stei:kim:2016}. Others have pushed back, citing evidence that it is rare to find situations when it is not preferable to condition on an additional pre-treatment covariate \citep{ding:mira:2014}.} However appropriately conditioning on many covariates either requires stronger parametric assumptions about the relationship between these potential confounders and the response variable or a sufficiently flexible approach to fitting this model. This tension has motivated a veritable explosion in the development of semi-parametric and nonparametric causal inference methodology in the past three decades.  How should applied researchers who rely on these tools choose among them?  This paper explores a new approach for comparing a wide variety of methods across a broad range of ``testing grounds'': a causal inference data analysis competition.  Overall we find strong evidence that approaches that flexibly model the response surface dominate the other methods with regard to performance.

\section{Motivation for causal inference competition}

Methodology for causal inference has been developed for a wide range of applications and leverages diverse modeling and computational techniques. The richness of the literature in this area---while offering numerous options to researchers---can also make it challenging to identify the most useful methodology for the task at hand.  There are several issues that further complicate this choice.

\subsection{Shortcomings of existing literature that compares performance of causal inference methods}

While many papers have been written in the past few decades proposing innovative technology, there are shortcomings to this medium as a way of providing information to researchers in the field about what method will be best for them. Strong performance of a method in a paper written by its inventor is encouraging but should be interpreted cautiously for the reasons discussed in this section.

\subsubsection*{Few methods compared and unfair comparisons.} 
Authors of causal inference methods papers most often compare their method to just a few competitors.  Typically comparisons are made to more traditional, and thus perhaps less ``cutting edge,'' methods. Moreover, even when more sophisticated competitors are included in the mix we suspect that, despite the best of intentions, these papers are still likely to be biased towards showing better performance for the method being introduced for several reasons.  

First, the authors of such papers are likely more knowledgeable about their own method than the competitors included in the empirical comparisons.  For example, an author might compare his/her proposed method to na{\"i}ve implementation of a method \citep[see, for instance,][]{hill:2011} or to an ``off-the-shelf'' version of a model that requires careful manipulation of tuning parameters for optimal performance. Second, authors may use metrics to evaluate methods that inadvertently bias their results towards favoring the method they propose. For instance they might focus on bias rather than root mean squared error or they may ignore confidence interval coverage.  

\subsubsection*{Testing grounds not calibrated to ``real life.''}
Methodological papers often compare the performance of a newly proposed approach to existing methods in the context of simulated data sets.\footnote{Real (that is, not simulated) observational data sets are sometimes used to motivate the issue but cannot point to a winner in the case of disparate findings.}  These approaches often test just a few different types of data generating mechanisms. Moreover attempts are not always made to calibrate these simulations to data that researchers typically encounter in practice, with a mix of types of variables (continuous, categorical, binary) and originating from joint distributions that may not be easily defined in a simulation paradigm.  For instance observed data from real studies are not likely to follow a multivariate normal distribution, though this distribution is often used to create simulated testing grounds.\footnote{A compromise position uses real data for covariates or possibly the treatment and then simulates the rest of the data \citep[outcome and possibly treatment assignment; for example see][]{hill:2011}.  Another compromise models the response surface using a highly saturated parametric model with many interactions and polynomial terms and uses that to simulate outcome data \citep*[for example see][]{kern:etal:2016}.}

On the other end of the spectrum, some researchers who develop methods are quite justifiably motivated by the inference problems they encounter while collaborating with subject-area experts. Consequently, simulations designed to test these methods may be very highly-tuned to mimic real life data but in a highly-specialized setting.  While it is natural to make sure that a method works in the specific scenarios for which it is designed, this doesn't necessarily help a general researcher understand how it might perform more broadly.

Some papers focus instead on the theoretical, typically asymptotic, properties of their method.  However, it may be equally difficult to map these mathematical properties to a given data set ``in the trenches'' where the sample sizes may be smaller than those required by the theory or other regularity conditions and distributional assumptions may not hold.

Yet another option for data to use as a testing ground are constructed observational studies \citep[see, for example,][]{lalo:mayn:1987,hill:reit:zanu:2004,shadish2008can}. These are studies that capitalize on data from both a randomized experiment and from an observational study or survey with similar participants and measures collected from a similar time period.  Constructed observational studies work by replacing the randomized control group from the experiment with the observational data and checking to see if it is possible to use observational study design or analysis methods to get an estimate of the treatment effect that is similar to the experimental benchmark.  While these studies, by construction, have the advantage of being highly calibrated to real world studies, they have several disadvantages.  First, each represents only one type of data generating process (DGP) (though possibly with minor variations if multiple comparison groups are used). Second, we never know in these studies whether ignorability is satisfied for the constructed study; therefore if an observational method is not able to recover the experimental treatment effect we cannot ascertain whether this is because ignorability was not satisfied or the model fit the data poorly.  Finally, since the comparison is between two \emph{estimates} it is not clear how to assess whether the observational estimate is ``close enough'' to the experimental benchmark.

\subsubsection*{File drawer effect.}
Understanding of the relative performance of methods can be biased due to the ``file drawer effect'' \citep{rosenthal1979file}.  Researchers searching for the best method for their problem won't have access to information on inconclusive comparisons since such results are unlikely to have been published. 

\subsection{Attempts to address these shortcomings through our competition}

The ``2016 Atlantic Causal Inference Conference Competition'' (henceforth referred to simply as the ``competition'') was initiated as an attempt to address the limitations of the current literature.  We discuss each concern outlined above in the context of this competition. 

\subsubsection*{Few methods compared and unfair comparisons.} 
The primary goal of the competition was to combat the issues of few and unfair comparisons.  First, we had a broad call and received submissions for 30 different methods.\footnote{We actually received a few more submissions than that however we only present results for methods from submitters who were willing to provide a description of the method. In addition there were two nearly-identical BART submissions so we only evaluate one of those and count only one towards this total.}  Each method was submitted by a researcher or team of researchers who we assume is knowledgeable about that method. As discussed more below, competition participants either implemented the method themselves or submitted a black box version of the method that they felt would work across the range of settings to be tested. In either case we assume that those submitting were sufficiently invested in that method's success to submit a competent implementation of the method.
Furthermore, we analyze method performance by considering a variety of performance metrics across a wide array of data features.\footnote{This approach is similar to a Kaggle competition \citep{carpenter2011may} though there was no public leaderboard or repeat submission opportunities provided; these can induce better performance \citep{athanasopoulos2011value} but also lead to overfitting to the test data \citep{wind2014model}. Indeed, there have been several prominent examples of crowdsourcing scientific tools \citep{vanschoren2014openml,ranard2014crowdsourcing,paulhamus2012crowdsourced}.}

Finally, the creators of this causal competition come from diverse fields and are accustomed to different types of data structures and norms regarding the acceptance of structural and parametric assumptions. This informs aspects of the simulation such as the magnitude of effects relative to unexplained variability, prevalence of nonlinearities or high-order interactions, numbers and kinds of covariates deemed reasonable to satisfy ignorability, average size of treatment effects, and range of biases in misspecified models. This should reduce the potential for the choice of type of data structure to favor one type of method over another.

A few other competitions have been run in the causal inference area.  For instance Guyon has organized several competitions for determining which of several candidate features was causally linked with an outcome \citep{guyon2008design}; these have focused on finding the causes of effects or determining temporal ordering of variables. As far as we can tell, our competition is the first with the more common statistical focus of estimating the effect of a cause in observational studies across a range of complications with regard to data features.

\subsubsection*{Testing grounds not calibrated to real life.}
As explained in more detail later in the paper, our simulations are built on covariates selected from a real data set.  Therefore this part of the data is calibrated to ``real life.'' In fact we chose a group of covariates that we thought might plausibly be included in a hypothetical study of the effect of birth weight on IQ -- this helped to mimic the types of natural correlations between covariates typical in an observational study.  We then simulated data from a wide variety of DGPs that reflect features of the data thought to be important for causal inference estimation: degree of nonlinearity, overlap/imbalance, percent treated, alignment between assignment mechanism and response surface, and treatment effect heterogeneity.  We hope the breadth of these simulation settings will at least have meaningful overlap with the breadth of observational data used in causal analyses occurring in practice today.

\subsubsection*{File drawer effect.}
Perhaps the most potent antidote to this potential problem is that we have published the code we used to create the simulations and evaluate the results on GitHub at \url{https://github.com/vdorie/aciccomp/tree/master/2016}. Therefore, anyone can test whatever method they want on the data using the evaluation metric of their choice.

\section{Notation and Assumptions}

We consider the causal effect of binary treatment $Z$, with $Z=0$ indicating assignment to control and $Z=1$ indicating assignment to treatment. $Y_i(0)$ is the outcome that would manifest for person $i$ if $Z_i = 0$; $Y_i(1)$ is the outcome that would manifest for person $i$ if $Z_i = 1$. Individual-level causal effects are defined as the difference between these ``potential outcomes,'' for example $Y_i(1) - Y_i(0)$ \citep{rubi:1978}. The observed outcome is defined as $Y = (1-Z_i) Y_i(0) + Z_i Y_i(1) $.

\subsection{Estimands}

Research often focuses on average causal effects across subpopulations of convenience or interest. We can formalize the average treatment effect as  $\E[Y(1) - Y(0)] = \E[Y(1)] - \E[Y(0)]$.  When this expectation is taken over the analysis sample this estimand is referred to as the sample average treatment effect (SATE).  Popular variants of this estimand restrict it by averaging over only those in the treatment group or conversely the control group to obtain, respectively, the sample average effect of the treatment on the treated (SATT) or the sample average effect of the treatment on the controls (SATC). Analogs of these treatment effects exist for the full population however these will not be addressed in this competition for reasons discussed below.

\subsection{Structural Assumptions}

Unfortunately we can never directly observe $Y(1)$ for observations assigned to control or $Y(0)$ for observations assigned to treatment. Thus these treatment effects are not identified without further assumptions. The most common assumption invoked to identify these effects is the so-called ignorability assumption \citep{rubi:1978}, which is also known as ``selection on observables,'' ``all confounders measured,'' ``exchangeability,'' the ``conditional independence assumption,''  and ``no hidden bias'' \citep[see][]{barn:cain:gold:1980,gree:robi:1986,lech:2001,rose:2002b}. A special case of the ignorability assumption occurs in a completely randomized experiment in which $Y(0), Y(1) \perp Z$. This property implies $\E[Y(a) \mid Z = a] = \E[Y(a)]$, which allows for identification of the above estimands solely from observed outcomes.

Observational studies typically rely on the more general form of the ignorability assumption, $Y(0), Y(1) \perp Z  \mid X$.  This allows for independence between potential outcomes and the treatment indicator conditional on a vector of covariates, $X$.  Thus identification can be achieved because $\E[Y(a) \mid Z = a, X] = \E[Y(a) \mid X]$. An average treatment effect can then be unbiasedly estimated by averaging the conditional expectation $\E[Y(1) - Y(0) \mid X] = \E[Y(1) \mid Z = 1, X] - \E[Y(0) \mid Z = 0, X]$ over the distribution of $X$. To obtain the ATT (or ATC), this averaging is performed over the distribution of $X$ for the treatment (or control) group. Although ignorability is typically an unsatisfying assumption to have to make, in the absence of randomized experiments or other environmental or structural conditions that give rise to various types of natural experiments (regression discontinuity designs, instrumental variables) few options are left.

Typically researchers invoke an even stronger assumption referred to as ``strong ignorability.''  This adds an assumption of overlap or ``common support.''  Formally this requires that $0 < \Pr(Z=1\mid X) < 1$ for all $X$ in the sample.  If this fails to hold then we may have neighborhoods of the confounder space where there are treated but no controls units or vice-versa.  That is, empirical counterfactuals  \citep{hill:su:2013} may not exist for all observations. Since many causal inference methods rely on some sort of modeling of the response surface, failure to satisfy this assumption forces stronger reliance on the parametric assumptions of the response surface model.

In the language of causal graphs \citep{pearl2009causality,pearl2009causal}, we can say that in our data generating process, the set of observed covariates $X$ form an admissible back-door adjustment set from the outcome $Y$ to the treatment $Z$. Therefore, by our construction, the causal effect is identifiable and can be estimated from the observed covariates $X$. We assume throughout that these covariates represent variables measured pre-treatment or that could not be affected by the treatment.  This leaves the problem of which statistical method is best suited for this task, which is the main focus of the challenge and of this paper.

\subsection{Parametric assumptions}

Even if we can assume ignorability, estimating causal effects without bias still requires estimating expectations such as $\E[Y(1) \mid X]$ and $\E[Y(0) \mid X]$. Estimating these conditional expectations can be nontrivial, especially in high dimensions, which is why there has been such a strong focus in the causal inference literature in the past few decades on appropriate ways to do so without making strong parametric assumptions \citep{kurt:etal:2006, hill:2011}.  This competition has a strong focus on how to achieve reliable and unbiased causal inference in an observational setting where the parametric assumptions may be difficult to satisfy.

\section{Testing grounds: Data and Generative Models}

Our goal was to generate data sets that are both useful in distinguishing between methods but that also exhibit the types of features typical of data from real studies. We describe our specific choices regarding creating the testing grounds and the logistics of the competition in this section.

\subsection{Embedded assumptions and design choices}

We imposed a small set of assumptions on all DGPs in the competition to make the competition practical and not overly complex.  

\subsubsection*{Ignorability.} 
We assumed ignorability throughout.  Approaches to nonignorable assignment of the treatment need to rely on an understanding of the context and the science of the problem.  Given that we are simulating our treatment and outcome we would either have had to invent ``science'' for how they were related to the covariates and each other or else map them to an existing scientific theory.  The former would require submitters to guess at our invented science.  The latter would require a great deal of subject matter expertise and which could unfairly bias some teams and theories over others. 
\subsubsection*{Estimand.} We needed to specify a causal estimand for researchers to estimate. We chose the effect of the treatment on the treated because most causal inference methods can easily target this estimand whereas a few do not naturally target the average treatment effect across the entire sample or population.  We focused on the \emph{sample} average treatment effect for the treated \citep[SATT;][]{hartman2015sample} because quite a few popular methods (for example matching) lack natural variance estimators for population estimands \citep{abad:imbe:2006}.

\subsubsection*{Overlap for the inferential group.}
Given our focus on SATT, to satisfy the overlap assumption we only need to ensure that empirical counterfactuals exist for all \emph{treated} units.  
When the estimand of interest is SATT and overlap \emph{does not} exist for the treatment group, many researchers would opt to reduce or reweight the inferential group (that is, the set of treated observations about which we will make inferences) to those for whom overlap is satisfied \citep{hira:imbe:2001, hira:imbe:ridd:2003, crum:imbe:etal:2009, hill:su:2013}. However, these approaches change the causal estimand. Given that it seems fair to allow for such a change in the causal estimand and yet some estimands are inherently easier to estimate than others, we decided that it would be too complicated to make the competition fair if we included settings in which common support for the inferential group was not satisfied.  

\subsubsection*{Simulating treatment and outcome.} To help calibrate our simulations to the type of data that might be analyzed by an empirical researcher we used covariates from a real study (discussed below).  We decided to simulate both the treatment assignment and the response due to two considerations. First, it allowed us to easily satisfy the ignorability assumption.  Second, it allowed us to manipulate specific features of the data such as balance, overlap, and nonlinearity of the model in ways that are directly measurable.  This enabled  exploration of the relationship between features of the data and method performance.

\subsection{Calibration to ``real data''}

Using an existing, real-world data set allowed us to incorporate plausible variable types as well as natural associations between covariates into the simulation.  We used data from the Collaborative Perinatal Project \citep{nisw:1972:cpp}, a massive longitudinal study that was conducted on pregnant women and their children between 1959 to 1974 with the aim of identifying causal factors leading to developmental disorders. The publicly available data contains records of over 55,000 pregnancies each with over 6,500 variables.

Variables were selected by considering a subset that might have been chosen for a plausible observational study. Given the nature of the data set we chose to consider a hypothetical twins study examining the impact of birth weight on a child's IQ. We chose covariates that a researcher might have considered to be confounders for that research question.  After reducing the data set to complete cases, 4802 observations and 58 covariates remained. Of these covariates, 3 are categorical, 5 are binary, 27 are count data, and the remaining 23 are continuous.

\subsection{Simulation procedure and ``knobs''}
\label{subsec:sim_proc}

We posit a data generating process (DGP) for the potential outcomes and the treatment assignment conditional on the covariates that factors their joint distribution as $p(Y(1), Y(0), Z | X) = p(Y(1), Y(0) \mid X) p(Z | X)$. Henceforth we refer to $p(Y(1), Y(0) \mid X)$ as the \emph{response surface} and $p(Z | X)$ as the \emph{assignment mechanism}. This factorization reflects our assumption of an ignorable treatment assignment, because in it $p(Y(1), Y(0) \mid Z, X) = p(Y(1), Y(0) \mid X)$. 

The assignment mechanism and response surface were generated according to a number of tunable parameters. Both models consisted of ``generalized additive functions,'' in which the contribution of covariates was first passed through a transformation function and then added or multiplied together. An example of such a function that includes two covariates is $f(x_i) = f_1(x_{i1}) + f_2(x_{i2}) + f_3(x_{i1})f_4(x_{i2})$, where $x_{ik}$ represents the $k^{\text{th}}$ covariate for the $i^{\text{th}}$ individual and $x_i$ is the vector of all such covariates for that individual. Here each $f_j$ might consist a sum of polynomial terms, indicator functions, or step functions. Moreover the sum of these terms could subsequently be passed through a ``link'' function, as in $g(x_i) = \exp(f_3(x_{i1}) + f_4(x_{i2}) + f_3(x_{i1})f_4(x_{i2}))$, permitting the output to be bounded or highly nonlinear.

Since the functions are generated randomly, the set of parameters that control the simulation framework essentially define a model over DGPs.  We refer to specific values of these parameters as simulation ``knobs''.  These knobs were adjusted to create simulation scenarios that produced structured deviations from idealized experiments to meaningfully test causal inference methods.  All combinations of knobs yield 216 scenarios.  Concern about the ability to test multiple replications in that many scenarios led us to focus on the most interesting 77 combinations by eliminating cases that were trivial or redundant, such as simple linear models or unconfounded treatment and response. Contest participants were told that there were 77 different scenarios, but that only 20 data sets would be used for the do-it-yourself portion of the competition. They were also told that there was a continuous outcome, a binary treatment indicator, and 58 covariates. Finally, they were informed that ignorability held throughout, that the observations in any given data set were identically and independently distributed, and that not all covariates were confounders.  For each of the 77 black-box scenarios, 100 independent replications were created, yielding 7700 different realizations.

The competition call described the simulation knobs 
as:  1) degree of nonlinearity, 2) percentage of treated, 3) overlap, 4) alignment, 5) treatment effect heterogeneity, and 6) magnitude of the treatment effect. Full details of the simulation framework can be found in Appendix~\ref{sec:appendix:simulation}; We provide an overview here (link to the R package that reproduces these datasets provided above).

After generating the data sets we also created a set of variables describing them, which are divided into what we call ``oracle'' and ``non-oracle'' variables, based on whether they would be available to a researcher with real-world data. We give examples of these metrics below, and provide a full list in Appendix \ref{sec:appendix:measures}.

\subsubsection*{Degree of nonlinearity.}
In the absence of nonlinear response surfaces and assuming ignorability, trivial estimation strategies such as ordinary linear regression are able to produce unbiased causal estimates. We are more interested in scenarios where the simplest methods fail \citep[many authors discuss the problems that occur with nonlinear response surfaces, including][]{imbe:2004, hill:su:2013,gelm:hill:2007, fell:2009}. Therefore we include nonlinearities in both the response surface and the assignment mechanism. In both cases, the nonlinearity is introduced by including higher order moments of individual covariates, interactions between covariates, and non-additive functions of the covariates. We allow the functional form of the covariates (the $f_j(x)$ terms from above) to include up to three-way-interactions and third-order polynomial terms, three-way-interactions and step functions, or remain linear. In general we restrict ourselves to additive models, however, when simulating response surfaces we optionally include a term that exponentiates a linear combination of covariates through the $g_k(x)$ term. An example of a variable that measures the degree of non-linearity in a given data set is Pearson's $R^2$ when regressing the observed outcome $Y$ on the observed, non-transformed covariates $X$. Across the $7700$ different realization this metric ranges between $0.02$ and $0.93$, with quartiles, $[0.26, 0.37, 0.48]$.
 
\subsubsection*{Percentage of treated.} Given that the estimand of interest is the effect of the treatment on the treated, any estimation approach might be challenged by having a low percentage of controls relative to the treated \citep{abad:imbe:2006}. We had two basic settings for this knob.  In one setting the expected value for the percentage of treated was 35\%, and in the other setting the expected value was 65\%.  In the low-treatment setting, 95\% of simulations had percentages of treatment between 0.20 and 0.38, while in the high-treatment setting 95\% of simulations were between 0.41 and 0.67.  The difference in ranges is a result of the overlap setting discussed next.

\subsubsection*{Overlap for the treatment group.} 
Despite deciding that overlap would be enforced for the treatment observations in all simulated data sets, we still wanted to explore the impact of having controls that are dissimilar from all treated units with regard to confounders.  This type of lack of overlap can be particularly challenging for methods that rely on models since many methods will attempt to extrapolate beyond the range of common support.
 
Thus we created low overlap settings in which we selected a ``corner'' of the covariate space and forcibly prevented observations with extreme values on several variables from receiving the treatment, regardless of whether they had a high propensity score; that is, we forced the propensity score for these observations to be zero. The more complicated the definition of this neighborhood, the more difficult it is for any method to identify the neighborhood as one that is fundamentally different from those where overlap exists and to avoid unwarranted extrapolation.  We then included the covariate interactions in the response surface to ensure alignment on them (see alignment discussion below).

One way we measure overlap is by calculating mean Mahalanobis distance between nearest neighbors with opposite treatment in the ground-truth covariate space (that is, the space in which the covariates have been transformed by the polynomial expansions used in the true assignment mechanism and response surface). This yields an oracle metric, since in general a practitioner will not have access to the true polynomial expansions. The quartiles of this metric for the cases where the \emph{overlap} knob was set to $1$ are $[3.73, 4.31, 4.99]$; the quartiles of the oracle metric for the cases where the \emph{overlap} knob was set to $0$ are $[4.16, 4.75, 5.50]$.
  
\emph{Balance}, defined as equality of covariate distributions across treatment groups, is a related concept to overlap since lack of overlap always implies lack of balance.  However imbalance can exist even when treatment and control groups have perfect overlap.  While we did not specifically target imbalance as a knob, we made sure that imbalance was achieved as a by-product of our specification of the treatment assignment mechanism.

A simple oracle metric of imbalance is the Euclidean norm of the distance between the mean of the control units and the mean of the treated unit in the ground-truth covariate space.  On this metric our simulation settings varied, with quartiles of $[0.78,    1.30,    2.68]$.

\subsubsection*{Alignment.}
The only covariates that have the potential to cause bias in our estimation of treatment effects are those that play a role in \emph{both} the assignment mechanism and the response surface.  For instance, when a covariate enters into the response surface or but not the assignment mechanism (or vice-versa), including that covariate in the estimation of the treatment effect may increase that estimator's efficiency but should not impact the bias.  Moreover, the functional form of the covariate is important. For instance, suppose age enters linearly only into the DGP of the assignment mechanism, however, it enters both linearly and with a squared term into the DGP of the response surface.\footnote{When we say that a covariate enters linearly into the model for the assignment mechanism we mean that it enters linearly into the part of the model equated to the inverse logit of the propensity score.}  If we include age linearly in our estimation strategy we should remove the bias that would be incurred by excluding it; from a bias perspective we do not need to account for the squared term.  However if the squared term is included in the data generating process of \emph{both} the assignment mechanism and the response surface, then we need to condition on the squared term our estimation strategy. Following \citet{kern:etal:2016} we refer to the correspondence between the assignment mechanism and response surface as \emph{alignment}.
Another framing of this issue is that the degree of alignment reflects the dimension of the confounder space.
  
The degree of alignment has several implications.  The first implication is that we can create more or less potential for bias conditional on an initial set of covariates by creating more or less ``alignment'' between the assignment mechanism and the response surface.  The second is that if the number of true confounders is small relative to the number of available covariates it may be difficult for a method to sort out which variables are the most important to privilege.  The third is that approaches that differentially privilege covariates that are strong predictors of either the treatment assignment or the response, but not both, may be at a disadvantage relative to those that are able target the true confounders.  

We varied the amount of alignment by altering the frequency with which terms appeared in both models.  This allowed us to create scenarios in which each model was highly complex, but only a specific fraction of terms in each model was a confounding term.  An oracle metric we used to reflect the degree of alignment is the correlation between the logit of the true assignment score $p(Z \mid X)$ and the outcome $Y$. Based on this metric the degree of alignment varied widely.  The absolute value of this correlation ranged from near 0 to about .94 with a median at approximately .29.
  
\subsubsection*{Treatment effect heterogeneity.}
There is no reason to believe that any given treatment affects all observations in the same way.  However parallel response surfaces, which yield constant treatment effects, are easier to fit than nonparallel response surfaces.  Creating heterogeneous treatment effects, or, equivalently, departures from parallel response surfaces adds to the computational and statistical challenges for causal inference methods. Treatment effect heterogeneity was created in our simulations by allowing certain terms in the response surface to have a different coefficient for $\E[Y(1) \mid X]$ than for $\E[Y(0) \mid X]$.  

An oracle measure used to capture this heterogeneity is the standard deviation of the treatment effect function $\E[Y(1) - Y(0) \mid X]$ across units within a setting, normalized by the standard deviation of the outcome within the same setting.  This metric varies across settings from $0$ to $2.06$ with quartiles at $[0.47, 0.73, 1.01]$.  Since treatment effects are all represented in standard deviation units with respect to the outcome measure, this represents a considerable amount of variability in the amount of heterogeneity that is present in any given data set.  For 200 realizations with \emph{treatment heterogeneity} knob set to $0$ the standard deviation is exactly $0$. 

\subsubsection*{Overall magnitude of the treatment effect.}  
Conference participants were alerted that the magnitude of the SATT would vary across settings.  While this was true by default we did not explicitly have a knob that directly manipulated the magnitude of the treatment effect.  Rather the other knobs implicitly created variation in this magnitude. The median SATT across 7700 realizations was 0.68 while the interquartile range stretched from 0.57 to 0.79 (again these are in standard deviation units with respect to the outcome).

\subsection{Issues not addressed}

As the first competition of its kind, we limited the scope of the problems addressed. We hope the authors of future competitions will find creative ways to explore some of the issues, described below, that we did not.

\paragraph{Non-binary treatment.}
Binary treatments are common in real studies though by no means the only type of causal variable of interest.  With regard to ignorability and overlap, binary treatments have the advantage of weaker and more transparent assumptions; two potential outcomes are easier to conceptualize and create overlap for than many potential outcomes.  Moreover,  in the absence of a linear relationship between treatment and outcome it can be difficult to find a simple characterization of the treatment effect.
  
\paragraph{Non-continuous response.} 
Disciplines vary dramatically in the extent to which their typical response variables are continuous or not.  Test scores and other continuous measures are common in fields like education, however political science and medicine often focus on binary outcomes (vote, survival).  Non-continuous responses can complicate causal inference because typical models have parameters that are not collapsible \citep{gree:robi:pear:1999}. That is, the marginal and conditional expectations from such models are not equal.
  
\paragraph{Non-IID data.}
Data with heterogeneous errors or correlations between responses are common. A method that cannot be generalized beyond the assumption of independent and identically distributed data has severely limited viability.
  
\paragraph{Varying data size or number of covariates.} 
Varying the number of observations, the number of covariates, or the ratio of the two has the potential to strongly affect the performance of methods.
    
\paragraph{Covariate measurement error.}
Measurement error in the covariates can lead to biased estimates of the treatment effect.  Few if any standard causal inference methods routinely accommodate this complication.  

\paragraph{Weakening the underlying assumptions of this competition.}
In addition to the issues above, future competition organizers might consider ways to violate our key assumptions of ignorability and overlap for the inferential group.

\subsection{Competition logistics}

This data analysis challenge was announced to the mailing list of the Atlantic Causal Inference Conference of about 800 people and on the conference website on April 21, 2016 (\url{http://jenniferhill7.wixsite.com/acic-2016/competition}).  Links were also distributed to several machine learning listserves and posted on a widely-read statistics blog that attracts readers from diverse disciplines. 

\section{Causal Inference Submissions and Key Features}

We received 15 submissions for the DIY portion of competition and 15 for the black-box section.  Two of the DIY submissions were not adequately described by the submitters and are thus omitted.\footnote{The organizers were not allowed to submit methods to the competition; however for the black-box competition we included a simple linear model using main effects to create a baseline for comparison (as distinct from the ``created'' methods described later).}

\subsection{Features of causal inference methods}

The submitted methods differed substantially in their approaches.   However, across the corpus of approaches to causal inference in observational studies there are a handful of features that have emerged as useful for distinguishing between methods.  We outline these and then use them to create a taxonomy by which to classify methods in Table \ref{tab:method_features}.  Additional details are provided in Appendix \ref{sec:appendix:methods}.

\subsubsection*{Stratification, matching, weighting.}
A strong focus of the causal inference literature over the past few decades has been on pre-processing data to reduce reliance on parametric assumptions \citep{scha:rotn:robi:1999, bang2005doubly, sekh:2007}.  Stratification, matching, and weighting all attempt to create treatment and control groups with similar covariate distributions. If sufficient balance can be achieved then estimation of the treatment effect can either proceed without a model or if a model is used, the estimate from the model should be fairly robust to misspecification.

In its purest form \emph{stratification} (or subclassification) creates balance by restricting comparisons of outcomes between treatment and control groups within the same cell of the contingency table defined by all of the covariates \citep{rose:rubi:1984}.  Variants of this have been proposed that stratify instead within leaves of a regression tree \citep{athey2015recursive, wager2015estimation}.

\emph{Matching} handpicks a comparison group for the treated by choosing only those control observations that are closest with regard to a given distance metric; comparison units that are not similar enough to the treated are dropped from the analysis \citep{stuart:2010}.The most popular distance metric currently is the propensity score although other choices exist \citep{rubin2006matched}.

\emph{Weighting} for causal inference is very similar to the type of weighting typically performed in the survey sampling world \citep{little1988missing}.  Similar to matching, the goal of weighting is to create a pseudo-population of controls that have a joint distribution of covariates that is similar to the joint distribution of covariates for the inferential group \citep*{rose:1987,robi:1999b}.  Thus controls are reweighted to look like treated observations when estimating the ATT, or vice-versa when estimating the ATC \citep{imbe:2004,kurt:etal:2006}.  To estimate the ATE both groups can be reweighted to reflect the covariate distribution of the full sample.  

\subsubsection*{Modeling of the assignment mechanism.}
Many methods that work to reduce reliance on parametric assumptions require accurate modeling of the treatment assignment mechanism, often because they incorporate the propensity score \citep{rose:rubi:1983}. The propensity score, defined as the probability that an individual receives the treatment given its covariates, $\Pr(Z \mid X)$, serves as a balancing score. Within the class of observations with the same balancing score, treatment assignment is ignorable. 

The propensity score is typically incorporated into a causal analysis through stratification, matching, or weighting. For instance one can stratify based on values of the propensity score, use the difference in propensity scores as the distance metric in matching, or weight on functions of the propensity score. The propensity score is also an important component of the Targeted Maximum Likelihood Estimation \citep[TMLE;][]{vand:robi:2003,  vand:rubi:2006} approach used in several of the submitted methods. TMLE is an approach to more efficiently estimating the causal effect and can be particularly helpful in situations with high-dimensional nuisance parameters.

\subsubsection*{Modeling of the response surface.}
If ignorability holds and sufficiently balanced treatment and control groups have been created using stratification, matching, or weighting then a simple difference in mean outcomes across groups will provide an unbiased estimate of the treatment effect. A parallel argument, however, is that if the response surface is modeled correctly it is not necessary to pre-process the data in this way \citep{hahn:1998, hill:2011}. Many causal inference approaches model both the assignment mechanism and the response surface.  While some approaches submitted fall either formally or informally under a ``double robust'' classification \citep*{robi:rotn:2001, vand:robi:2003} we prefer to categorize methods separately by whether they model the assignment mechanism or the response surface since it is beyond the scope of our efforts to decide whether each method ``formally'' qualifies as doubly robust.

\subsubsection*{Nonparametric modeling of the assignment mechanism or response surface.}
Almost every modern-day causal inference approach involves either modeling of the assignment mechanism (typically, although not always, to estimate the propensity score) or modeling of the response surface.  However, flexible modeling of the assignment mechanism was largely ignored until the past decade, particularly in the context of propensity score estimation, because it was seen by many primarily as a means to an end for achieving good balance \citep{lee2010improving, westreich2010propensity}.  If subsequent matched, re-weighted, or subclassified samples failed to achieve a given threshold for balance the model was typically modified by, for instance, adding or removing interactions and quadratic terms, or performing  transformations of inputs until adequate balance was achieved.  In recent years, however, more attention has been given to estimation of propensity scores using methods that require less strict parametric assumptions than traditional methods such as logistic or probit regression \citep[e.g.][]{westreich2010propensity}.

Nonparametric modeling of the response surface has also received some attention over the past decade \citep{hill:2011, wager2015estimation, taddy2016nonparametric}. This is an alternative to approaches that try to create balanced samples so that estimation methods are robust to misspecification of the response surface.   

\subsubsection*{Variable selection.}
Researchers often have access to far more potential confounders than are realistic to include in any given analysis.  It can be helpful to exclude variables that are not true confounders. Variable selection techniques such as LASSO \citep{tibshirani1996regression} and the elastic net \citep{zou2005regularization} can help reduce the scope of the estimation problem to true confounders, so that more complicated algorithms can be used on the terms that really matter.\footnote{Some of the submissions are based on machine learning techniques that implicitly perform variable down-weighting or selection.  We do not label the methods in this ``gray area'' as variable selection methods in this taxonomy.}

\subsubsection*{Ensemble methods.}
It is unrealistic to expect that any one causal inference method can to perform well across all potential settings.  {\em Ensemble} methods mitigate this concern by running each of several methods on a data set.  Relative performance of the methods is evaluated using cross-validation or model averaging.  Then either the estimate from the best method is chosen or estimates from several methods are combined in a weighted average where the weights are based on relative performance \citep{dietterich2000ensemble}.\footnote{While some definitions of ensemble methods might include Random Forests or even methods that include boosting such as BART as ensemble methods, we use a more narrow definition that requires a library of competing methods that are all fit \emph{separately} to the same data and where the methods are weighted using performance metrics, where the weights might be 0's and 1's. \citet{rokach2009taxonomy} reviews this distinction.}

\subsection{Overview of submissions}

\newcolumntype{C}{>{\centering\let\newline\\\arraybackslash\hspace{0pt}}m{0.42cm}}
\newcommand{\tbp}[1]{\parbox[t]{0.41cm}{\bfseries #1}}
\newcommand{\rbb}[1]{\rotatebox{270}{{\scriptsize \bfseries \parbox[t]{3.45cm}{\hspace*{-0.25cm} #1 \hspace*{\fill}}}}}

\begin{table}[ht!]
\begin{center}
  \rowcolors{2}{lightgray}{}
  {\small \begin{tabular}{r|CCCCCCCCCC}
  \toprule
  \bfseries \parbox[t]{3cm}{Method Name} & \tbp{ST} & \tbp{MC} & \tbp{WT} & \tbp{PS} & \tbp{PS\\NP} & \tbp{PS\\VS} & \tbp{RS} & \tbp{RS\\NP} & \tbp{RS\\VS} & \tbp{EN} \\
  \midrule
  Ad Hoc &  &  &  &  &  &  & X &  & X &  \\ 
  Bayes LM &  &  &  &  &  &  & X &  &  &  \\ 
  Calibrated IPW &  &  & X & X &  &  &  &  &  &  \\ 
  DR w/GBM+MDIA 1 &  &  & X & X & X &  & X & X &  &  \\ 
  DR w/GBM+MDIA 2 &  &  & X & X & X &  & X & X &  &  \\ 
  IPTW estimator &  &  & X & X &  & X & X &  &  &  \\ 
  GLM-Boost &  &  & X & X &  &  &  &  &  &  \\ 
  LAS Gen Gam &  & X &  &  &  & X & X & X & X &  \\ 
  Manual &  & X &  & X &  &  & X &  &  &  \\ 
  MITSS & X &  &  & X &  &  & X & X &  &  \\ 
  ProxMatch &  & X &  &  &  &  &  &  &  &  \\ 
  RBD TwoStepLM & X &  &  & X &  &  & X &  &  &  \\ 
  Regression Trees &  &  & X & X & X &  & X & X &  &  \\ 
  VarSel NN &  & X & X & X & X & X & X &  & X &  \\ 
  Weighted GP &  &  &  &  &  &  & X & X &  &  \\ \hline
  Adj. Tree Strat & X &  &  & X & X &  & X &  &  &  \\ 
  Balance Boost &  &  & X & X & X &  &  &  &  &  \\ 
  BART &  &  &  &  &  &  & X & X &  &  \\ 
  calCause &  &  &  &  &  &  & X & X &  & X \\ 
  CBPS & X &  &  & X &  &  & X &  &  &  \\ 
  h2o Ensemble &  &  & X & X & X &  & X & X &  & X \\ 
  LASSO+CBPS &  &  & X & X &  & X & X &  & X &  \\ 
  Linear Model &  &  &  &  &  &  & X &  &  &  \\ 
  MHE Algorithm &  &  &  &  &  &  & X &  &  &  \\ 
  SL + TMLE &  &  & X & X & X & X & X & X & X & X \\ 
  teffects ra &  &  &  &  &  &  & X &  &  &  \\ 
  teffects ipw &  &  & X & X &  &  &  &  &  &  \\ 
  teffects ipwra &  &  & X & X &  &  & X &  &  &  \\ 
  teffects psmatch &  & X &  & X &  &  &  &  &  &  \\ 
  Tree Strat & X &  &  & X & X &  & X &  &  &  \\ \hline
  \rowcolor{white} & \rbb{Stratification} & \rbb{Matching} & \rbb{Weighting} & \rbb{Propensity Score Fit} & \rbb{PS Nonparametric} & \rbb{PS Variable Selection} & \rbb{Response Surface Fit} & \rbb{RS Nonparametric} & \rbb{RS Variable Selection} & \rbb{Ensemble} \\
  \end{tabular}}
  \end{center}
  \caption[Method Summary]{Summary of methods. First block are do-it-yourself methods and second block are black-box. Within blocks, methods are in alphabetical order.}
  \label{tab:method_features}
\end{table}

Table~\ref{tab:method_features} summarizes the methods submitted in terms of the above features. Many submissions involved novel or complex approaches. Most explicitly estimated a propensity score. Most fit a model to the response surface.  More than half used some sort of weighting.  Matching was almost entirely absent from the black-box portion of the competition, while techniques like variable selection were scattered throughout. The black box methods and their submitters favor sophisticated nonparametric modeling techniques.
 
The success of the competition and academic value in understanding its results owe entirely to the large number of high quality submissions from researchers in this field. We are incredibly grateful to all researchers who submitted something to our competition in the hopes of furthering discussion on methodology and best practices for making causal inferences from observational data. A full list of participants is in Appendix~\ref{sec:appendix:acknowledgements}.

\subsection{Top performers}

The results of the competition are fully elaborated in section~\ref{sec:results}. In this section we describe in a bit more detail the five methods among the original submissions that had the best performance overall.

\subsubsection*{Bayesian Additive Regression Trees (BART).}
Bayesian Additive Regression Trees (BART), is a nonparametric method for fitting arbitrary functions using the sum of the fit from many small regression trees.  This black box submission incorporates priors over the tree structure and tree predictions to avoid overfitting \citep*{chip:geor:mccu:2010}. The BART method for causal inference fits the joint function $f(x, z)$ which is then used to draw from the posterior predictive distribution for both $y(1) = f(x, 1)$ and $y(0) = f(x,0)$.  The empirical posterior distribution for any given average effect can be obtained by taking the differences between these quantities for each person at each draw and then averaging over the the observations about which we want to make inferences \citep{hill:2011}.

\subsubsection*{Super Learner plus Targeted Maximum Likelihood Estimation (SL+TMLE).}
This black box submission was an ensemble algorithm. Super Learner uses its library of methods to make out-of-sample predictions through cross-validation, which are then combined according to weights that minimize the squared-error loss from predictions to observations. The weights are used to combine the fitted values from the methods when fit to the complete data set.  This approach then applies a TMLE correction to Super Learner estimates.  

The ensemble library consisted of {\tt glm}, {\tt gbm}, {\tt gam}, {\tt glmnet}, and {\tt splines} (all functions in {\tt R}) to model both assignment mechanism and response surface \citep{poll:etal:2016}. The fit from assignment model was incorporated into the response surface using propensity score weights by expanding the population average treatment effect on the treated into conditional average treatment effects across individuals \citep{hira:imbe:ridd:2003}.

\subsubsection*{calCause.}
The calCause black box submission was an ensemble algorithm that uses cross-validation to chose between Random Forests \citep{brei:2001} and a Gaussian process \citep{rasm:will:2006} with an unknown covariance function for fitting the response surface of the controls. The method with better out-of-sample prediction was used to impute the control response for the treated observations, which were differenced from the \emph{observed} treated responses and then averaged. Uncertainty was measured by bootstrap resampling.

\subsubsection*{h2o.}
This black box submission labeled h2o refers to the open source deep learning platform, h2o.ai \citep{h2o:2016}. As implemented here, it is an ensemble approach that performs ``super learning'' separately for the assignment mechanism (to predict the propensity score) and the response surface for controls (to predict $Y(0)$ for the treated observations). Observed outcome values for the treated were used to predict $Y(1)$.  Differences in predicted counterfactuals were averaged using IPTW ATT weights. Models in the ensemble library include: {\tt glm}, {\tt random forest}, {\tt deep learning (NN)}, {\tt LASSO}, and {\tt ridge regressions} \citep{lede:2016}.
  
\subsubsection*{DR w/GBM + MDIA 1 and 2}
This DIY submission used generalized boosted regression to estimate separate models for the assignment mechanism and response surface, each allowing for up to three-way interactions. The model for the assignment mechanism was used to estimate treatment-on-treated weights for control cases. These weights were then calibrated using Minimum Discriminant Information Adjustment \citep[MDIA;][]{habe:1984}\footnote{MDIA is also referred to alternately in other literatures as ``calibration weighting,'' ``exponential tilting'' or ``entropy balancing''.  Each creates weights of a simple exponential form that minimally perturb of a set of base weights (in this case, ATT weights) to exactly meet pre-specified constraints.  In this case, the approach started with ATT weights for the control cases and then calibrated them so that the weighted mean of each covariate, and a new covariate equal to the estimated control mean function, was exactly equal to the corresponding treatment group mean.} to achieve exact balance of the means of both the individual covariates and the estimated response surface. It is equivalent to a type of bias-corrected doubly-robust estimator.  Confidence intervals were constructed using bootstrap standard errors.  The first submission (DR w/GBM + MDIA 1) used 5-fold cross validation to select both the response and selection models.  The second submission (DR w/GBM + MDIA 2) used 50,000 trees for both models, which tended to be many more trees than cross-validation would choose, particularly for the selection model.

\subsection{Post-competition methods}

After the preliminary results of the competition were presented at the Atlantic Causal Inference Conference in May 2016 (http://jenniferhill7.wixsite.com/acic-2016/competition), the competition was re-opened for new submissions.  The Super Learner team availed themselves of this option and submitted a revised version of their original Super Learner submission that included BART in the library (SL+BART+TMLE). Moreover, we found that an important means for exploring  which features of the most competitive methods were most important for success was by tweaking and combining some of the originally submitted methods.  These investigations resulted in the nine additional methods discussed now.

The two top-performing ensemble-based submissions modeled both the assignment mechanism and the response surface. To explore the contribution from modeling both mechanisms rather than just the response surface alone, we created an augmented version of the BART submission that also modeled the assignment mechanism (BART IPTW).  Since the default version of BART is known to sometimes perform poorly when predicting binary outcomes \citep*{dori:hara:carn:hill:2016}, cross-validation was used to choose the parameters for the prior distributions in the BART fit for the assignment mechanism.  As with SL+TMLE and calCause, the fit from the assignment mechanism was incorporated into the BART estimate through IPTW weights.

To explore the role of the TMLE adjustment in the superior Super Learner + TMLE performance, we re-ran that algorithm without TMLE correction (Super Learner).  Relatedly, we also augmented the original BART submission with IPTW plus the TMLE correction (BART+TMLE), where the propensity score is again fit using BART with cross-validation.

Since BART was the only stand-alone method to rival the performance of the ensemble methods, we performed further tweaks to the originally submitted approach.  First, we used a BART fit just to the response surface using cross-validation to choose the hyperparameters rather than using the default prior (BART Xval).  Another new BART approach used several chains from distinct starting points (typically BART is run with one chain); this was also fit just to the response surface (BART MChains).  A third BART approach altered the multiple chains approach simply to report symmetric intervals rather than percentile intervals (MBART Symint).  Finally, we implemented a version of BART where the estimated propensity score is included as a covariate (BART on PScore); this was inspired by work by \citet{hahn:etal:2017} that develops extensions of BART that focus on estimation of treatment effect heterogeneity less prone to the biases that can be caused by regularization.

One difference between the SL and BART approaches is that the original SL submission fits separate reponse surface models for the treatment and control conditions while the BART submission considers treatment as a covariate and fits the response surfaces jointly.  This motivated creation of a method that re-wrote the SL approach to fit the response surfaces for the treated and control observations simultaneously (SL+TMLE Joint). The specific ensemble in this implementation included BART.

\section{Evaluation of Performance of Submitted and Constructed Methods}
\label{sec:results}

We assess all submitted methods across the 20 DIY data sets. We evaluate both the BB competition submissions and the post-competition methods across the 7700 BB methods.  This section describes our global summaries of performance for each of the two competitions. 

We evaluate root mean squared error (RMSE) to get a better sense of how close on average each treatment effect estimate is to the true value of the estimand. Bias was calculated as the average distance between the causal estimate and estimand (SATT) across all data sets.  When reporting for the black box methods we additionally display the interquartile range of all biases across the 77 settings and 100 replications.  

Interval coverage reflects the percentage over all data sets that the reported interval covers the true SATT.  Given the potential trade-offs between coverage rates and interval length we also report the average interval length.  A less conventional measure of performance is the PEHE (precision in estimation of heterogeneous effects) \citep{hill:2011}. Within a given data set this measure reflects the root-mean-squared distance between individual level treatment effect estimates and individual level differences in true potential outcomes. We report the average PEHE across all data sets.  Since the DIY competition only had 20 methods we felt that coverage measures and PEHE would be too imprecise to be useful so we do not report these measures for those methods in the body of the paper.  We do report these in Appendix \ref{sec:appendix:extra}.    Finally, computational time was calculated for the black box submissions (we could not observe this for the DIY methods).

\subsection{Comparison of all methods in the 20 DIY data sets}

\begin{figure}[h]
  \begin{center}
    \includegraphics[height=3.5in,width=4in]{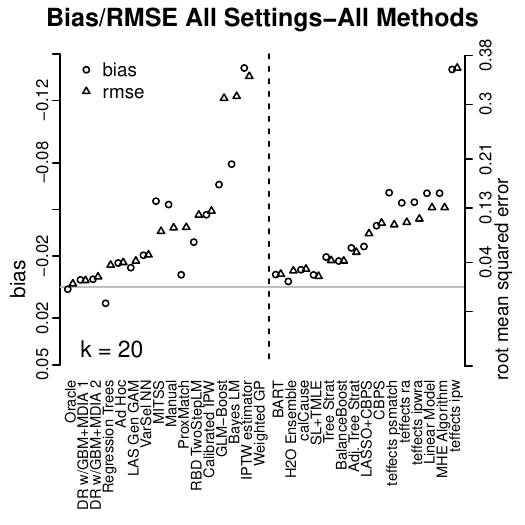}
  \end{center}
  \caption{This plot displays both bias, represented by circles (values on the left y-axis), and RMSE represented by triangles (values on the right y-axis).  Both are calculated across the 20 data sets in the DIY competition.  The dashed vertical lines divides the DIY submissions (to the left) from the black box submissions (to the right).}
  \label{fig_biasrmse_diy}
\end{figure}

All of the submissions in both the DIY and black box competitions were run on the 20 DIY data sets.  Figure~\ref{fig_biasrmse_diy} displays the results with respect to RMSE and bias.
The DIY submissions are on the left side of the vertical dividing line in the plot; the black box submissions are to the right of the line.  Within the first group the submissions are ordered by performance with respect to RMSE, while within the second they are ordered with respect to their RMSE in the complete black box set of simulations.  The ``oracle'' results correspond to performance when the conditional average treatment effect on the treated is used as the ``estimate.''\footnote{The conditional average treatment effect on the treated is defined as $\E[Y(1)-Y(0) \mid X, Z=1]$ where the expectation represents the average over the treated units in the sample.}

Given that the performance of the DIY submissions can only be evaluated over 20 data sets, we are reluctant to draw strong conclusions about relative performance.  However based on bias and RMSE we see strongest performance across \emph{all} methods from the two DR w/GBM+MDIA submissions.  The top black box performers (discussed above) are the next best performers.  It is worth noting that the DIY methods that perform next best all rely on flexible fitting of either the assignment mechanism, the response surface, or both.  Additional results are provided in Appendix \ref{sec:appendix:extra}.

\subsection{Comparison of black box submissions}
We now compare all the methods submitted to the black box competition with regard to all metrics described above across all 7700 data sets.  

\begin{figure}[t]
  \begin{center}
    \includegraphics[height=3.5in,width=4in]{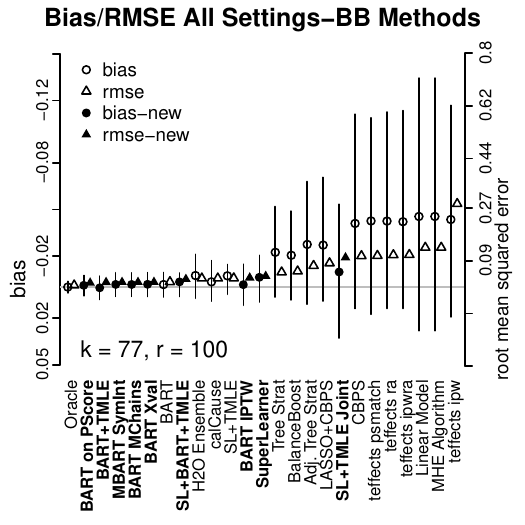}
  \end{center}
  \caption{This plot displays both bias (left y-axis) and RMSE (right y-axis) for all submitted black box methods and newly created methods.  Both are calculated across the 7700 data sets in the black box competition.  Bias is displayed by circles and RMSE by triangles, each averaged across all the data sets; open symbols are used for submitted methods and filled for newly created methods.  Lines for bias measures show the interquartile range of all biases across the 77 settings and 100 replications.}
  \label{fig_biasrmse_bb}
\end{figure}

\subsubsection*{RMSE and bias.}
Figure \ref{fig_biasrmse_bb} displays the results with regard to RMSE (triangles, scale on right y-axis) and bias (circles, scale on left y-axis).\footnote{A noteworthy feature of these results is that the average bias is negative for all methods.  This reflects the fact that the treatment effect distribution had a positive expected value and most methods shrink their treatment effect estimates towards zero.}  Methods are arrayed across the x-axis in order of performance with respect to RMSE.  Of the originally submitted methods, four stand out with respect to bias and root mean squared error:  BART, SL + TMLE, calCause, and h2o. The next most obvious group of methods with superior performance are Tree Strat, BalanceBoost, Adjusted Tree Strat and LASSO + CBPS.  In general it is fair to say that most of the methods in this competition performed reasonably well with both (the absolute value of) bias and RMSE at or below about .05 standard deviations with respect to the outcome measure; this level of standardized bias is very low relative to traditional measures of effect sizes \citep*{cohe:1962}.\footnote{Another way to think about the absolute level of the performance of the methods is to see how closely it resembles that of the oracle.}  In particular, and not surprisingly, the methods created or submitted after the initial deadline all performed particularly well relative to the others with the exception of SL+TMLE joint, which likely failed because many of the methods in that ensemble library were incapable of fitting non-parallel response surfaces and thus were unable to estimate heterogeneous treatment effects.

\begin{figure}[t]
  \begin{center}
    \includegraphics[height=3in,width=3.5in]{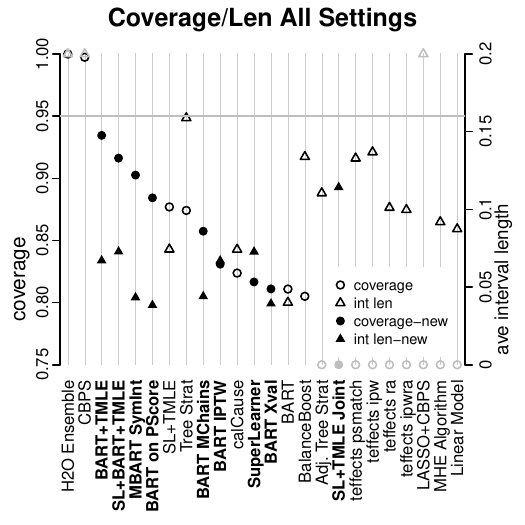}
  \end{center}
  \caption{Coverage (circles) and average interval length (triangles) for all of the black box and newly created methods across the 7700 black box data sets.  Methods are ordered according to decreasing coverage rates. Methods in bold/filled plot points represent the newly created methods.  Points in gray were beyond the plotting region (very poor coverage or very large intervals) and are shown at the corresponding top or bottom edge.}
  \label{fig_covlen_bb}
\end{figure}

The following additional comparisons however are perhaps of note.  Adding BART to the SL+TMLE submission improved performance relative to SL+TMLE without BART in the library, however this addition was not sufficient to outperform BART as a stand-alone method.\footnote{It is possible that this is because BART was used to fit the assignment mechanism without any adjustment to account for the fact that standard BART-for-binary implementations do not always perform well with binary outcomes \citep{dori:hara:carn:hill:2016}.} Three of the BART augmentations led to slightly better performance with respect to bias and RMSE (BART+TMLE, BART MChains, BART Xval) and one did not (BART IPTW). 

\subsubsection*{Interval coverage and length.}
While many of the automated algorithms submitted to this competition performed well with regard to RMSE and bias, performance varied widely with regard to interval coverage and length.  All of the originally submitted methods were somewhat disappointing in this regard.  Figure \ref{fig_covlen_bb} displays these results. This plot displays both coverage (circles) and average interval length (triangles) for all of the black box and newly created methods across the 7700 black box data sets.  The methods are ordered according to decreasing coverage rates.\footnote{The one triangle that sits on the 95\% line is potentially misleading; recall this point displays \emph{interval length}, not coverage.}  Since all intervals are intended to have 95\% coverage we plot a horizontal line at that level. The best methods will have coverage close to 95\% while also having the shortest intervals.  Methods in bold/filled plot points represent the newly created methods.  Points in gray were beyond the plotting region (very poor coverage or very large intervals) and are shown at the top or bottom edge The last nine methods on the right all had coverage below 75\%, while  CBPS and h2o Ensemble had average interval lengths of $0.78$ and $6.1$.

Several of the new methods were added to address concerns about confidence interval coverage.  These augmentations were successful to varying degrees. One successful augmentation with respect to coverage rates was BART + TMLE.  While the original BART implementation had average coverage around 82\%, the BART + TMLE implementation had nearly nominal coverage.  This was accompanied by approximately a 50\% increase in average interval length but this length is still slightly smaller than the other top-performing methods. The BART implementation that includes the propensity score also increases coverage noticeably to a little over 90\% without any increase in average interval length. Finally, the BART implementation that reports symmetric rather than percentile intervals (MBART SymInt) results in an increase in coverage rates that is similar with only a small increase in average interval length.  Another successful augmentation was the addition of BART to the SuperLearer + TMLE ensemble. This resulted in a shift in average coverage from about 83\% to about 92\% with no noticeable increase in the average interval length.

\subsubsection*{Precision in Estimation of Heterogeneous Effects.}
Only a subset of the methods output the individual level treatment effect estimates required to calculate the Precision in Estimation of Heterogeneous Effects (PEHE) measure\footnote{The TMLE and IPTW extensions of BART affect only how individual effects are averaged to calculate an estimate of the average treatment effect for the treated.} \citep{hill:2011}.  Of these, the BART methods and calCause performed noticeably better than the other options; however, since the two other competitors were quite simple and relied on linear models this was not particularly surprising.  Given the importance of targeting treatments based on covariate profiles it would be useful if future competitions continued to include this measure in their evaluations and encouraged submitters to submit individual level effect estimates in addition to an estimate of the average treatment effect for each data set.

\subsubsection*{Computational time.}
Another metric that discriminated sharply across the methods was computational time, measured as the system time of the process spawned for the method. As the cluster is a shared resource, run times for individual replications within a simulation setting varied widely. To compensate, average run time was computed by taking the median run time within each setting and then the arithmetic mean of those values. Even with this adjustment, usage varied widely from week to week so that the run time should be seen as a rough measure and not equivalent to the performance in a controlled environment.

One might expect that all of the ensembles would perform substantially worse on this metric.  However, h2o, at 24.8 seconds, was quite competitive with BART, at 29.4 seconds. It should be noted that the h2o method frequently failed to run and required numerous restarts, so that its times may be unreliable. The method also offloads much of its computation to background processes, so that the resources consumed are difficult to measure. Methods that use cross validation, including all of the Super Learner submissions and some of the modifications of the BART algorithm, took substantially longer.  Many of these still remained in the 10 to 15 minute range,  however three approaches were noticeably more computationally expensive: calCause (27.7 minutes), SuperLearner (46.9 minutes), and SL+TMLE Joint (46.8 minutes).

\section{Predicting Performance}

As discussed Section~\ref{sec:results}, certain methods outperform the majority of submissions. Thus one could simply advise researchers to use one of the top-performing methods. However, we wanted to be able to provide advice about which methods work best in specific types of settings. Moreover, we wanted to understand which characteristics of a method are associated with strong performance. We use the performance data from the black box competition to address these questions.\footnote{The DIY competition did not have enough data to perform similar types of analyses.}

Overall, we find that to a surprising degree we could not go beyond a general recommendation of using flexible non-parametric response surface modeling. Under almost no condition could we predict which of the black box methods would outperform others, beyond its average level of performance across all settings. In other words, to the degree that we can examine it in these testing grounds, relative performance is much less contextual than we had imagined.

\subsection{Measures used to predict performance}

We created a set of 25 metrics that describe the 77 different experimental settings and their instantiations. These include both levels of simulation ``knobs'' (described in \ref{subsec:sim_proc}), as well as metrics measured directly on the simulated data. These data-based metrics are divided into ``oracle'' metrics, which rely on information only available to those with knowledge of the true DGPs, and ``non-oracle'' metrics that are available directly from the observed data. An example of an oracle metric is the correlation between the true propensity score and the outcome; an example of a non-oracle metric is the pseudo-$R^2$ achieved when estimating a propensity score model using logistic regression. While oracle metrics provide a more accurate representation of the features of the data, non-oracle metrics could in principle be evaluated before selecting a method and thus guide the choice.  The full list of metrics is given in Appendix \ref{sec:appendix:measures}.

\subsection{Performance variance explained}

We first attempted to explain variation in performance for each method, one at a time. We built multiple linear regression models, one for each of the 24 black-box methods, explaining the log of the absolute bias for each of the 7700 data realizations. The amount of variation in performance is method dependent, but using the log transform reduces differences in the amount of targeted variance across methods.  The predictors in these models are the metrics used to describe the 7700 data realizations. Interpreting the log as relative change on the original scale, total variation is similar for all methods on the log scale, and thus $R^2$ for our explanatory models may be compared across methods.

Across methods the $R^2$ from the predictive models rarely exceeds 0.10 when predictive models include only non-oracle measures. Essentially, we cannot deduce, from data-derived metrics, how well a method will perform on a given dataset. When we add in oracle knowledge, this increases to 0.40-0.50 for just over a third of the methods in the competition, mostly the weaker-performing methods. For the more successful methods, even with the oracle knowledge the $R^2$ rarely exceeds 0.10. Details are provided in Appendix \ref{sec:appendix:explaining_variance}. Essentially, our ability to predict performance for a given method based on features of the data is very poor for most methods and is roughly inversely proportional to the overall performance of that method.  

\subsection{Cross-method analysis about features of data and models}

The above analysis is conditional on the method, and as such does not explain differences in performance \emph{across} methods, nor whether there are settings in which certain types of methods perform better than others. We evaluate these questions using a sequential set of multilevel models (MLM) (described in Appendix \ref{sec:appendix:explaining_variance}) that partition the variation into components reflecting differences in performance between methods, settings and their interaction, as well as ``unexplained'' variation due to differences between realizations net of all else. 

This partitioning estimates that 37\% of the variation is attributable to differences in the average performance of methods (see Table \ref{tab:mlm_vc} and discussion in Appendix \ref{sec:appendix:explaining_variance} for details). Using dummy codings for the features of methods displayed in Table \ref{tab:method_features}, we can explain 76\% of the between method, average performance differences, net of all else. Inclusion of a non-parametric fit to the response surface accounts for most of these differences.

Variance attributable to settings, at 5\%, is a small portion of the total variation in performance, but we are able to explain essentially all of these differences, on average, using two non-oracle data features. The first is a measure of {\em nonlinearity} of the response surface: the $R^2$ of the observed outcome $Y$ on the observed design matrix. The second is a measure of the degree of {\em alignment} between the assignment mechanism and response surface: the $R^2$ between the estimated unit level treatment effect estimated by BART  and propensity scores estimated with logistic regression on the observable design matrix. The data conditions that are almost completely predictive of poor performance are poor alignment between outcome and treatment assignment mechanisms and non-linearity of the response surface. We were unable to find important method by data condition interactions.  

In summary, whether we use oracle or non-oracle data measures, features of methods, or the interactions between these, we remain exceedingly limited in our capacity to explain differential performance at the realization level.  The unexplained across dataset variation in performance still accounts for over half of the total variation.

\section{Discussion}

We have created the first large-scale data analysis competition for estimating causal effects in the context of observational studies. Through our efforts to summarize and unpack the results several themes emerged.  
One theme is that of all the ways we created complexity in the data, two of the three that created most the most difficulty across the board for achieving low bias were nonlinear response surfaces and treatment effect heterogeneity. Perhaps not surprisingly then, methods that were able to flexibly model the response surface routinely distinguished themselves as high performers. This held true even for methods like BART that only modeled the response surface and not the assignment mechanism.  Moreover these methods had superior performance relative to approaches that only focused on flexibly modeling of the assignment mechanism (Balance Boost, Tree Strat, Adj. Tree Strat).  This also helps to explain the superior performance of most of the ensemble methods which were able to capitalize on the relative strengths of a variety of models in order to achieve this flexibility.

Another theme has to do with the fact that lack of alignment across the assignment mechanism and the response surface emerged as one of the most challenging features of the data.  This data feature has been discussed only rarely in the causal inference literature \citep[see, for instance,][]{kern:etal:2016}. Lack of alignment creates difficulty because if there are many covariates available to a researcher and only a subset of these are true confounders (and indeed perhaps only certain transformations of these act as true confounders) then methods that are not able to accurately privilege true confounders are potentially at a disadvantage.  Of course most of the submissions did not explicitly do this. However quite a few approaches performed either explicit variable selection or implicit weighting of predictors based on some metric of importance.\footnote{It can be argued that almost all regression type methods perform this kind of weighting of course.  However they do so using different metrics and to different degrees.}  It appears that there is a bigger payoff to this type of selection or re-weighting of inputs in the response surface modeling however.  This is consistent with advice warning against variable selection in the assignment mechanism as well as advice to focus attention on the relative importance of inputs to the response surface \citep*{aust:groo:2007b,hill:2008, pear:2010}.

The third theme is that good coverage was difficult for most methods to achieve even when bias was low.  While we were able to achieve better coverage by ``tweaking'' some of the best-performing methods (in particular the TMLE adjustment often seemed beneficial, though it did not uniformly improve coverage), we don't feel like we have strong advice about how to optimize this aspect of performance.

On the positive side, a final theme is that there are \emph{several} good options for accurately estimating causal effects, particularly if the primary goal is to reduce bias; furthermore, many of these have R packages that are readily available.  Of course that advice comes with the caveat that our testing grounds have been restricted a range of settings where ignorability holds, overlap for the inferential group is satisfied, the data are i.i.d., etc.; these properties may not hold in practice and far more work needs to be done to understand what methods may work in settings with additional complications.  

This competition has provided a vehicle for evaluating the efficacy of a wide range of methods across a much broader set of DGPs for the assignment mechanism and response surface than is typically present in any single methodological paper. These comparisons were bolstered by the ability to ``crowdsource'' the methodological implementations.  We hope that our efforts will inspire others to create similar types of data analysis competitions that explore different types of challenges so that we can continue to learn together as a community about how to create methods that will reliably perform well for applied researchers across a wide range of settings.

\bibliographystyle{imsart-nameyear}
\bibliography{competition,causal}

\appendix

\section{Appendix section}
\label{sec:appendix}

\subsection{Details of Simulation Framework}
\label{sec:appendix:simulation}

We described above the knobs that were manipulated to create the 77 different simulation settings.  This appendix specifies the levels of each knob that are used for each of the black box and DIY data sets.  Here we provide more information below about the knob settings.  The R package at \url{https://github.com/vdorie/aciccomp/tree/master/2016} recreates the simulated data by setting parameters according to the following enumeration.

{\small
\begin{longtable}{rlllllll}
\label{tab:simulation_settings} \kill
\caption[Simulation Settings]{Simulation Settings} \\
  \toprule
      & {\bf Treatment} & {\bf Trt} && {\bf Response} & \parbox[t]{1.75 cm}{\bf Trt/Rsp} & {\bf hetero-} & {\bf DIY} \\
    {\bf \#} & {\bf Model} & {\bf \%} & {\bf Overlap} & {\bf Model} & \parbox[t]{1.75 cm}{\bf Alignment} & {\bf geneity} & {\bf \#} \\
  \midrule
  1 & linear & low & penalize & linear & high & high & 10 \\ 
  2 & polynomial & low & penalize & exponential & high & none & 1 \\ 
  3 & linear & low & penalize & linear & high & none & 9 \\ 
  4 & polynomial & low & full & exponential & high & high & 4 \\ 
  5 & linear & low & penalize & exponential & high & high & 15 \\ 
  6 & polynomial & low & penalize & linear & high & high & 2 \\ 
  7 & polynomial & low & penalize & exponential & high & high & 5 \\ 
  8 & polynomial & low & penalize & exponential & none & high & 13 \\ 
  9 & step & low & penalize & step & high & high & 8 \\ 
  10 & linear & low & penalize & exponential & low & high & 14 \\ 
  11 & polynomial & low & penalize & linear & low & high & 19 \\ 
  12 & polynomial & low & penalize & exponential & low & high & 12 \\ 
  13 & linear & high & penalize & exponential & high & high & 18 \\ 
  14 & polynomial & high & penalize & linear & high & high & 20 \\ 
  15 & polynomial & high & penalize & exponential & high & high & 6 \\ 
  16 & polynomial & high & penalize & exponential & none & high & 17 \\ 
  17 & step & high & penalize & step & high & high & 7 \\ 
  18 & linear & high & penalize & exponential & low & high & 3 \\ 
  19 & polynomial & high & penalize & linear & low & high & 16 \\ 
  20 & polynomial & high & penalize & exponential & low & high & 11 \\ 
  21 & polynomial & low & penalize & step & low & low &  \\ 
  22 & polynomial & low & penalize & step & low & high &  \\ 
  23 & polynomial & low & penalize & step & high & low &  \\ 
  24 & polynomial & low & penalize & step & high & high &  \\ 
  25 & polynomial & low & penalize & exponential & low & low &  \\ 
  26 & polynomial & low & penalize & exponential & high & low &  \\ 
  27 & polynomial & low & full & step & low & low &  \\ 
  28 & polynomial & low & full & step & low & high &  \\ 
  29 & polynomial & low & full & step & high & low &  \\ 
  30 & polynomial & low & full & step & high & high &  \\ 
  31 & polynomial & low & full & exponential & low & low &  \\ 
  32 & polynomial & low & full & exponential & low & high &  \\ 
  33 & polynomial & low & full & exponential & high & low &  \\ 
  34 & polynomial & high & penalize & step & low & low &  \\ 
  35 & polynomial & high & penalize & step & low & high &  \\ 
  36 & polynomial & high & penalize & step & high & low &  \\ 
  37 & polynomial & high & penalize & step & high & high &  \\ 
  38 & polynomial & high & penalize & exponential & low & low &  \\ 
  39 & polynomial & high & penalize & exponential & high & low &  \\ 
  40 & polynomial & high & full & step & low & low &  \\ 
  41 & polynomial & high & full & step & low & high &  \\ 
  42 & polynomial & high & full & step & high & low &  \\ 
  43 & polynomial & high & full & step & high & high &  \\ 
  44 & polynomial & high & full & exponential & low & low &  \\ 
  45 & polynomial & high & full & exponential & low & high &  \\ 
  46 & polynomial & high & full & exponential & high & low &  \\ 
  47 & polynomial & high & full & exponential & high & high &  \\ 
  48 & step & low & penalize & step & low & low &  \\ 
  49 & step & low & penalize & step & low & high &  \\ 
  50 & step & low & penalize & step & high & low &  \\ 
  51 & step & low & penalize & exponential & low & low &  \\ 
  52 & step & low & penalize & exponential & low & high &  \\ 
  53 & step & low & penalize & exponential & high & low &  \\ 
  54 & step & low & penalize & exponential & high & high &  \\ 
  55 & step & low & full & step & low & low &  \\ 
  56 & step & low & full & step & low & high &  \\ 
  57 & step & low & full & step & high & low &  \\ 
  58 & step & low & full & step & high & high &  \\ 
  59 & step & low & full & exponential & low & low &  \\ 
  60 & step & low & full & exponential & low & high &  \\ 
  61 & step & low & full & exponential & high & low &  \\ 
  62 & step & low & full & exponential & high & high &  \\ 
  63 & step & high & penalize & step & low & low &  \\ 
  64 & step & high & penalize & step & low & high &  \\ 
  65 & step & high & penalize & step & high & low &  \\ 
  66 & step & high & penalize & exponential & low & low &  \\ 
  67 & step & high & penalize & exponential & low & high &  \\ 
  68 & step & high & penalize & exponential & high & low &  \\ 
  69 & step & high & penalize & exponential & high & high &  \\ 
  70 & step & high & full & step & low & low &  \\ 
  71 & step & high & full & step & low & high &  \\ 
  72 & step & high & full & step & high & low &  \\ 
  73 & step & high & full & step & high & high &  \\ 
  74 & step & high & full & exponential & low & low &  \\ 
  75 & step & high & full & exponential & low & high &  \\ 
  76 & step & high & full & exponential & high & low &  \\ 
  77 & step & high & full & exponential & high & high &  \\   \bottomrule
\end{longtable}}

\begin{itemize}
  \item {\bf Treatment model} - determines the base library of functions used when building the treatment assignment mechanism, $P(Z = 1 \mid X)$. {\em linear} implies that some predictors $x_{.j}$ are added to the assignment mechanism model as linear terms with random coefficients, {\em polynomial} gives a chance that, for continuous predictors, quadratic or tertiary terms are added in addition to a ``main effect'', and {\em step} potentially adds ``jumps'' and ``kinks'' of the form $\mathrm{I}\{x \leq A\}(x_{.j})$ and $(x_{.k} - B)\mathrm{I}\{x \leq C\}(x_{.k})$ respectively.
  \item {\bf Trt \%} - the baseline percentage of observations receiving the treatment or control conditional. Ranges from 35\% to 65\%.
  \item {\bf Overlap} - when not {\em full}, a {\em penalty} term was added to the linear form of the treatment assignment mechanism ($\mathrm{logit} P(Z = 1 \mid X)$) that added a large, negative value for combinations of extreme values of randomly chosen covariates. That is, terms of the form $A \cdot I\{x_{.j} > B\} \cdot I\{x_{.k} \leq C\} \cdots$, where $A$ is large and $B$, $C$, ... chosen from marginal quantiles. The penalty term forcibly excludes some observations from the treated population.
  \item {\bf Response model} - similar to {\em trt model}, determines the library of functions used when building the response surface, $\E[Y \mid X, Z]$. {\em exponential} encompasses the {\em polynomial} condition but adds a single term of the form $\exp\{f_1(x_{.j}) + f_2(x_{.k})\}$, with sub-functions that are linear, quadratic, or step.
  \item {\bf Trt/Rsp alignment} - achieved by specifying a marginal probability that a term in the treatment assignment mechanism is copied to the response surface. {\em low} gives an approximate 25\% chance, while {\em high} gives an approximate 75\% one.
  \item {\bf Heterogeneity} - controls the number of terms with which treatment interacts. \em{none} implies that treatment is a single, additive term in model, {\em low} implies that treatment is interacted with approximately three of the terms in the response model, and {\em high} yields around six interactions.
\end{itemize}

In treatment and response model generation, coefficients are generally drawn from Student-$t$ distributions when unbounded or beta-prime distributions when strictly positive. Assuming that covariates are scaled to approximately $[-1, 1]$, sub-function locations and scales are generated to approximately map to the same region. After functional terms are chosen, combined functions are rescaled so that the observed inputs yield plausible results, that is propensity scores almost exclusively in the range of 0.1-0.9 and response variables with a mean of 0 and expected standard deviation of 1. Finally, the response surfaces for the treated and controls ($Y(1)$ and $Y(0)$) are adjusted to meet a generated average treatment effect. The response surface noise and treatment effect are generated from heavy-tailed distributions.

\subsection{Glossary of Submitted and Created Methods}
\label{sec:appendix:methods}

The following table describes all of the methods considered in this paper whether they were competition submissions or created after the fact by the organizers.  In the case of competition submissions the descriptive wording was largely contributed by the person or team submitting the method for consideration, with some light editing for space.  In cases where the submitter failed to respond to requests for a description the methods was excluded from inclusion in the paper.  In all such cases the methods were low performers.  Methods are ordered alphabetically.

\renewcommand{\tbp}[1]{\parbox[t]{2cm}{#1}}
{\small
\begin{longtable}{lp{0.78\textwidth}}
\label{tab:diy_methods} \kill
\caption[Do-It-Yourself Methods]{Do-It-Yourself Methods} \\
  \toprule
  \tbp{\bfseries Method\\Name} & {\bfseries Description} \\
  \midrule
  Ad Hoc & The method first used GBM to screen for variables that predicted control group outcomes. It then made ad hoc decisions about variable transformations, and applied unmentionable and unreplicable incantations with the goal of improving fit and the ability to extrapolate beyond the range of the observed data.  It then used stepwise AIC to select among models allowing for up to three-way interactions among the subset of variables chosen from the previous steps. The selected model was then used to predict control outcome values for all treatment cases. It is not an automated method and has only tenuous grounding in statistical theory. \\
  Bayes LM & Naive Bayesian linear model. \\
  \tbp{Calibrated\\IPW} & Estimates a logistic propensity score calibrated such that the causal estimator is unbiased under the assumption that the response for the controls is linearly related to the covariates. Variance estimates are obtained from an asymptotic approximation. \\
  \tbp{DR w/GBM+\\MDIA} & The method used generalized boosted regression models (GBM) with cross validation to estimate flexible response and treatment models. The treatment model was used to obtain treatment-on-treated (TOT) weights, and these weights were tweaked with Minimum Discriminant Information Adjustment (MDIA) to achieve exact balance of the means of both the individual covariates and the estimated response surface. It is equivalent to a type of bias-corrected doubly-robust estimator. \\
  GLM-Boost & Boosted generalized linear model and bootstrapping for the confidence intervals. \\
  \tbp{IPTW\\estimator} & A stabilized inverse probability of treatment weighting method for the ATT. The propensity score is estimated by first selecting only covariates highly correlated with response. Weights are then used to regress response on treatment, and confidence intervals are obtained by bootstrap resampling. \\
  \tbp{LAS Gen\\GAM} & Least absolute shrinkage and selection operator (LASSO) was first used to select covariates for both treatment and response variables separately. Covariates selected for the treatment model were used in the GenMatch algorithm, which selects a yields a single control for each treatment. The responses for that constructed data set were fit using a generative additive model (GAM) using as predictors treatment and covariates from either variable selection model. \\
  Manual & Hand-done logistic regression followed by matching and weighted linear regression. \\
  MITSS & Multiple imputation with two subclassification splines (MITSS) explicitly views causal effect estimation as a missing data problem. Missing potential outcomes are imputed using an additive model that combines a penalized spline on the probability of being missing and linear model adjustments on all other orthogonalized covariates. \\
  ProxMatch & A matching method based on the proximity matrix of a random forest, in which treatment and control observations that tend to end up in the same terminal nodes are matched. \\
  \tbp{RBD\\TwoStepLM} & We first stratify on the estimated propensity score fitted by a linear model to approximate randomized block designs, then use the linear regression adjustment to analyze the randomized experiment within each stratum of the estimated propensity scores, and finally combine the estimates to get the overall estimator for the average treatment effect on the treated. \\
  \tbp{Regression\\Trees} & Bootstrap estimates of the propensity score were generated using decision trees; several models for estimating SATT, including individual trees and a boosted ensemble, then used sample weights based on these estimates. \\
  VarSel NN & Random Forest (RF) and LASSO variable selection steps were used on the response to determine variables for a neural network (NN) propensity score model. Non-overlaping treated observations were eliminated by a caliper distance, a matching set from the remaining made using the optmatch package, and from this ATT weights extracted. Finally, the response was regressed on treatment and selected variables with the aforementioned weights. \\
  \tbp{Weighted\\GP} & Uses a Gaussian process to model the response for the treatment group and a weighted Gaussian process to model the control group. Weights were derived so as to solve the ``covariate shifting'' problem, in which groups have different marginal distributions but the same conditional. This was done by applying the Frank-Wolfe optimization algorithm to the the Kullback-Leibler Importance Estimation Procedure, minimizing the KL divergence between the two distributions. Finally, predicted responses were averaged over treatment group to get estimates and confidence intervals. \\
  \bottomrule
\end{longtable}}

{\small
\begin{longtable}{lp{0.78\textwidth}}
\label{tab:bb_methods} \kill
\caption[Black Box Methods]{Black Box Methods} \\
  \toprule
  \tbp{\bfseries Method\\Name} & {\bfseries Description} \\
  \midrule
  \tbp{Balance-\\Boost} & A boosting algorithm is used to estimate the propensity scores by maximizing a covariate balancing score. An outcome regression is subsequently applied to adjust the estimate and estimate the maximum bias that might come from using an IPTW estimator. The bias and variance are combined to form conservative confidence intervals (wider and have more than 95\% coverage). \\
  BART & This approach uses a Bayesian nonparametric method (Bayesian Additive Regression Trees) to flexibly model the response surface. The method can produce posterior distributions for both average and individual-level treatment effects. \\
  \tbp{BART IPTW} & A joint BART model that uses cross-validation to choose a hyperparameter when fitting the assignment mechanism and the default parameters when fitting the response surface. The results are combined using an propensity-score weighted difference. \\
  \tbp{BART MChains} & BART fit only to the response surface using default settings, but combining the results of multiple chains. \\
  \tbp{BART on PScore} & BART MChains with a propensity score calculated using cross-validation, as in BART IPTW. Propensity score is added to the covariates in the response model. \\
  \tbp{BART + TMLE} & The joint model from BART IPTW but with the addition of the TMLE correction. \\
  \tbp{BART + Xval} & A response surface-only model that uses cross-validation to chose BART's hyperparameters. \\
  calCause & Response surface for controls fit by using cross validation to choose between random forests and a Gaussian process with a kernel matrix estimated using ``FastFood'' method. These imputed counterfactuals were paired with the observed treated values and bootstrap sampling used to obtain a standard error. \\
  CBPS & The propensity score was estimated by maximizing a balancing score (CBPS), which was then used to stratify observations. Independently, a linear model with third order polynomial terms was fit to the controls and used to make predictions for the treated. The ATT was then estimated by a weighted combination of the averages across strata. \\
  \tbp{h2o\\Ensemble} & Ensembler learners (glm, RF, ridge, deeplearner, ...) model response for controls and propensity score. These are then combined using IPTW ATT weights to take the difference between observed and predicted treated values. \\
  Linear Model & Linear model/ordinary least squares. \\
  \tbp{LASSO+\\CBPS} & This method estimated the SATT using propensity score reweighting. Propensity scores were estimated via the covariate balancing propensity score method proposed by Imai \& Ratkovic (2014), after selecting covariates with a preliminary LASSO regression that included main effects for each covariate. The model for the response surface was then selected using a weighted LASSO regression that included interaction terms and polynomial terms (for continuous covariates). \\
  \tbp{MBart SymInt} & BART MChains but with intervals calculated using a normal approximation instead of the empirical quantiles of the posterior samples. \\
  \tbp{MHE\\Algorithm} & The state of the art in labor and development econometrics: ordinary least squares and robust standard errors. \\
  SL+TMLE & Targeted minimum loss-based estimation (TMLE) was implemented using a universal least-favorable one-dimensional submodel. The outcome regression and propensity scores were modeled using super learning, with a library consisting of logistic regression models, gradient boosted machines, generalized additive models, and regression splines. Covariates supplied to the Super Learner were pre-screened based on their univariate association with the outcome. \\
  \tbp{SL + BART + TMLE} & The Super Learner/TMLE algorithm with BART added to the set of models. \\
  \tbp{Super \\ Learner} & The Super Learner algorithm without the TMLE correction. \\
  teffects ipw & Stata teffects function - inverse probability weighting using logistic regression and first order terms. \\
  teffects ipwra & Stata teffects function - ipw + logistic and weighted OLS. \\
  \tbp{teffects\\psmatch} & Stata teffects function - matching using logistic regression on first order terms and nearest neighbor. \\
  teffects ra & Stata teffects function - regression adjustment by fitting separate models to treatment and control with first order terms. \\
  Tree Strat & Tree-based stratification for average treatment effect estimation. The method first trains a CART tree to estimate treatment propensities, and then uses the leaves of this tree as strata for estimating the ATT. \\
  \tbp{(Adj.) Tree\\Strat} & An adjusted variant of the Tree Strat method, it seeks to improve the fit via a regularized regression adjustment in each stratum. \\
  \bottomrule
\end{longtable}}

\subsection{Extra DIY Results}
\label{sec:appendix:extra}

\begin{figure}[ht]
  \begin{center}
    \includegraphics[width=3.2in,height=2.6in]{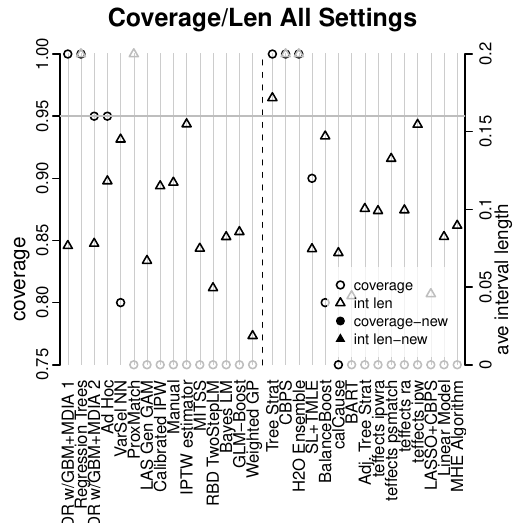}
   \end{center}
  \caption{This plot displays coverage (circles) and average interval length (triangles) for all of the DIY and original black box methods across the 20 DIY data sets. Methods are ordered according to decreasing coverage rates. Methods in bold/filled plot points represent the newly created methods. Points in gray were beyond the plotting region (very poor coverage or very large intervals) and are shown at the top or bottom edge.}
  \label{fig_diy_extra1}
\end{figure}

\begin{figure}[ht]
    \includegraphics[width=3.2in,height=2.4in]{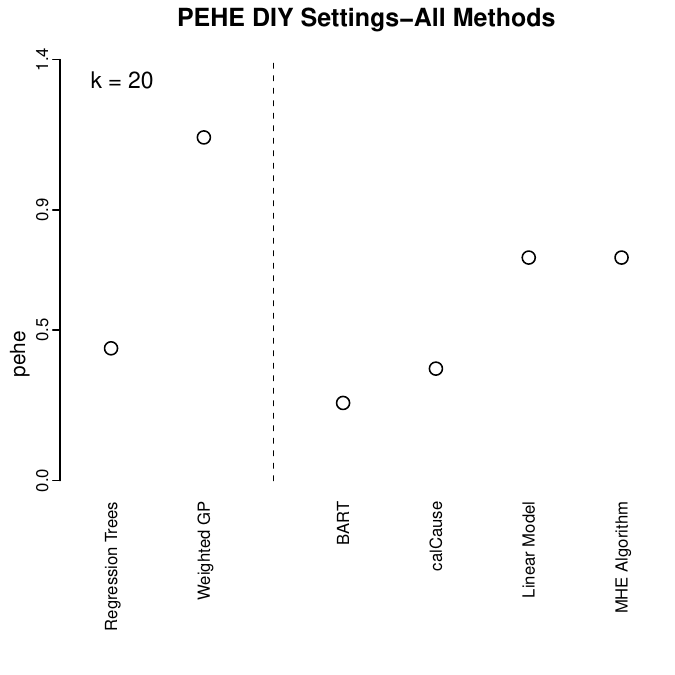}
   \caption{This plot displays PEHE for the DIY and black box methods that supplied individual-level treatment effect estimates.}
  \label{fig_diy_extra2}
\end{figure}

This section contains supplementary results for the Do-It-Yourself portion of the competition, including coverage, interval length, and precision in estimation of heterogeneous effects (PEHE). These are reported in Figures \ref{fig_diy_extra1} and \ref{fig_diy_extra2}.

\subsection{Explaining variance: modeling results}
\label{sec:appendix:explaining_variance}

Table \ref{tab:method_summary} displays the $R^2$ of the model for each method's performance in terms of $\log$ absolute bias linearly regressed on the 25 metrics describing the experimental conditions plus quadratic terms (see list in Appendix \ref{sec:appendix:measures}).  The table is ordered by the overall RMSE performance of the methods. The first column reports the $R^2$ from the regression utilizing the non-oracle metrics, and these tend to explain very little variance. The next column (2) evaluates a model with indicator variables for the 77 settings only; one could characterize this as modeling the average outcome per setting, and we find that a bit more variation is explainable with these, especially for the worst performing methods. Next, in column 3, labeled ``All Metrics,'' the regression is on the oracle and non-oracle measures, and these more fine-grained (realization level) measures do seem to be able to explain a bit more variation.  Lastly, column 4, ``Settings $+$ Metrics'' adds an indicator variable for each of the $77$ experimental conditions (as per column 2) to the metrics in column 3, resulting in a bit more variation explained.

\begin{table}[ht!]
  \rowcolors{2}{lightgray}{}
  {\small 
  \begin{tabular}{r|cccc}
  \toprule
  \bfseries \parbox[t]{3cm}{Method Name} & No Oracle & Only Settings  & All Metrics & Settings + Metrics\\
  \midrule
  teffects ipw & 0.09 & 0.23 & 0.37 & 0.38 \\ 
MHE Algorithm & 0.18 & 0.42 & 0.48 & 0.50 \\ 
    Linear Model & 0.17 & 0.41 & 0.47 & 0.49 \\ 
teffects ipwra & 0.09 & 0.27 & 0.43 & 0.45 \\ 
teffects ra & 0.09 & 0.29 & 0.44 & 0.46 \\ 
teffects psmatch & 0.07 & 0.22 & 0.33 & 0.35 \\ 
CBPS & 0.08 & 0.28 & 0.43 & 0.45 \\ 
SL+TMLE Joint & 0.27 & 0.17 & 0.39 & 0.40 \\ 
LASSO+CBPS & 0.11 & 0.22 & 0.36 & 0.38 \\ 
Adj. Tree Strat & 0.06 & 0.17 & 0.19 & 0.22 \\ 
BalanceBoost & 0.11 & 0.21 & 0.32 & 0.34 \\ 
Tree Strat & 0.01 & 0.14 & 0.13 & 0.16 \\ 
SuperLearner & 0.10 & 0.10 & 0.19 & 0.20 \\ 
SL+TMLE & 0.04 & 0.04 & 0.08 & 0.09 \\ 
BART IPTW & 0.06 & 0.10 & 0.21 & 0.22 \\ 
calCause & 0.03 & 0.08 & 0.09 & 0.10 \\ 
h2o Ensemble & 0.03 & 0.08 & 0.09 & 0.10 \\ 
SL+BART+TMLE & 0.03 & 0.03 & 0.06 & 0.06 \\ 
BART & 0.03 & 0.05 & 0.09 & 0.10 \\ 
BART Xval & 0.03 & 0.06 & 0.09 & 0.10 \\ 
MBART SymInt & 0.02 & 0.05 & 0.09 & 0.10 \\ 
BART MChains & 0.03 & 0.06 & 0.09 & 0.10 \\ 
BART+TMLE & 0.03 & 0.04 & 0.08 & 0.09 \\ 
BART on PScore & 0.02 & 0.04 & 0.09 & 0.10 \\ 
      \bottomrule
  \end{tabular}¬
 }
 \caption{Squared correlation coefficient of regressing each methods' performance in terms of $\log$ absolute bias, across $77$ experimental conditions times $100$ repetitions, onto the 25 metrics describing the experimental conditions plus quadratic terms (see list in Appendix \ref{sec:appendix:measures}), in the ``All Metrics'' setting. The ``Settings $+$ Metrics'' column is the same, with an additional dummy variable for each of the $77$ experimental conditions.}
  \label{tab:method_summary}
 \end{table}
 
We supplemented the separate regression models with a single multilevel model, again with log absolute bias as the outcome, and the metrics described previously as predictors.  By examining all methods together, we are able to partition the total variation into that associated with methods, settings or their interaction.  The prior analysis ignored the effect of methods and its potential interaction with settings.  Another way to understand the multilevel approach is that average performance for a particular set of conditions may be more easily predicted than performance of a single realization.  In fact, we organize the predictors in the multilevel model so that some of them are the average value of the metric across 100 realizations, yielding effectively 77 unique values per predictors per method. Including these in the model yields the implicit assessment of the average performance for each of the 77 setting scenarios.  It will turn out that we have the ability to predict \emph{average} performance across the 100 realizations {\em very well}.   

In our multilevel model, we include group random effects for twenty-four methods, and group random effects for the 77 conditions set by us for the contest. A simple unconditional means model with these effects  allows us to partition the variance in performance into components attributable to methods and settings.  As is common in this modeling paradigm, we then attempt to ``explain'' these variance components via the metrics described in Appendix \ref{sec:appendix:measures} and by features of the methods described in table \ref{tab:method_features}.  As in the method-specific analysis of the prior subsection, the former are divided into those estimable by the researcher (non-oracle) and those only known by those charged with data generation (oracle). In practice, the researcher is thus quite limited in terms of this information, so one of our objectives is to determine the extent to which this matters. 

The baseline model, given in column 1 of table \ref{tab:mlm_vc}, succinctly describes the sources of variation in this study.  Methods and Settings, along with their interaction, sum to 1.135, which is 46\% of the total variation.  The remaining 54\% of the variation labeled `Realizations' is at the trial level and reflects both the idiosyncratic error and the variability of the predictors and their interrelationships within setting.  Somewhat surprising is the smaller amount of variation captured in the interaction of settings and methods, 0.091, which suggests that methods are simply better or worse, and not dramatically better {\em than another method} in a given setting. In column 2, we add a set of eight non-redundant indicators for features of the method, as described in table \ref{tab:method_features}.  The indicators explain 76\% of the main effects for variation between methods, but only a few features are significant: utilizing a non-parametric response surface is by far the most important feature, improving the outcome dramatically, by a factor of -2 on the log scale.  

In columns 2 and 3, we add our oracle and non-oracle metrics after pre-processing them as follows: first, we rescale them to have mean zero and variance one; then, we compute the mean of these for each setting and center the metric within setting to reflect only its deviation; we include those setting-level means; and we include squared versions of each of those paired terms to allow for simple non-linearities. This form of centering within cluster \citep{enders2007centering} will isolate the impact of mean metrics to between group effects and centered metrics to within group effects.

In column 3, we see that the non-oracle metrics are able to explain a large portion of the main effects for variation between settings (81\%).  This suggests that the researcher may gain substantial knowledge of their setting from quantities derived from observables, but recall that the limited magnitude of variance associated with the setting-by-method interaction implies that this knowledge should not help much in choosing a method. Including the oracle metrics of settings, we can explain nearly all of the main effects for the corresponding variation.  Thus, if we have a good idea of what type of setting we were in with respect to alignment, non-linearity of the response surface, etc., we would have a decent sense of how well our method would perform in terms of (log) absolute bias, but the choice of method would probably remain fairly steady: a good method will use a non-parametric model of the response surface.

Our last two columns reflect our maximal explanatory power.  In column 5, we interact the metrics with characteristics of the method derived from table \ref{tab:method_features}, non-oracle first.  These should target the variance component associated with the setting by method interaction, but it could also explain realization variance.
For the non-oracle interactions, we explain 45\% of the setting by method variance component. We make little progress explaining the idiosyncratic (realization) variance with this set of interactions.  With the addition of oracle metrics,  as given in the last column, which reflect qualities of the settings without being exact values of the settings themselves, we explain 70\% of the interaction between setting and method, which is an improvement, but perhaps at a cost of requiring an unrealistic form of knowledge. By the end, we do make a little progress explaining 6\% of the realization variation.

We can conclude that realizations within setting are hard to explain, but most of the setting and method variation is not. Perhaps a bit surprising is how much we can learn from observables -- at least we can assess our likely performance in the given context even if it will be quite difficult to have this influence our choice of method.  Another conclusion is that it is possible to predict how a method will perform on average in a given contextual setting, but not much more; that is, ``explaining'' variance components involves explaining mean differences between groups. We know that we do well, on average, with methods that flexibly model the response surface, and we understand the conditions under which we should expect to do a bit less well.

The conditional regression analysis of the prior subsection differs from this multilevel analysis because it ignores the variation between methods expressed by row 1 in table \ref{tab:mlm_vc}, but we can verify the relationship between the approaches as follows. If we exclude row 1 from the total variation, we are then trying to explain 2.443-0.914 = 1.53 units of variation.  In column 6, we see that 1.475-0.217 = 1.26 units remain unexplained, suggesting that on average, we can explain about 18\% of the total variation (ignoring between method variation, but averaged across methods).  This is consistent with the results in table \ref{tab:method_summary}, in which the last column reveals a range of $R^2$ between 0.06 and 0.50 for the model that has a full set of indicators for the 77 settings along with all metrics.
 
\begin{table}[t]
\centering
\begin{tabular}{|l|c|c|c|c|c|c|} \hline
Variance & Uncond. & Method & \multicolumn{2}{c|}{Metrics} &  \multicolumn{2}{c|}{Feature $\times$ Metric} \\
Component & Mean & Features & Non-Oracle & +Oracle & Non-Oracle & +Oracle  \\ \hline
Methods             & 0.914 &  0.216 & 0.216 & 0.216 & 0.217 & 0.217 \\
Settings           &  0.130 & 0.130 & 0.025 & 0.002 & 0.027 & 0.005 \\
Setting $\times$ Methods & 0.091 & 0.091 & 0.091 & 0.091 & 0.050 & 0.027 \\
Realizations  & 1.308 & 1.308 &  1.288 & 1.269 & 1.272 & 1.225 \\ \hline
Total              & 2.443 & 1.744 &  1.620 & 1.578 & 1.566 & 1.475 \\ \hline
\end{tabular}
\vspace{.1in}
\caption{Variance components analysis:  Results from six different multilevel models predicting log absolute bias from the black box methods across the 7700 data sets.  The first column displays the partition of the variance in log absolute bias explained by the methods and the 77 settings relative to the unexplained variance across data set realizations within setting and method.  The other columns show how the variation explained by each component changes as we include features of the models, then non-oracle metrics, then oracle metrics, and the interactions between these sets of metrics (non-oracle and non-oracle plus oracle) and the method features.}
\label{tab:mlm_vc}
\end{table}

\subsection{Full list of metrics used to describe experimental settings}\label{sec:appendix:measures}

List of metrics used to describe the experiments. We denote $\left[\text{oracle}\right]$ those metrics which are not accessible to a researcher in an ordinary observational study, but are available to us as the creators of the competition data set. We further denote by \knob the oracle metrics that correspond to the explicit experimental settings we created, as described in Subsection \ref{subsec:sim_proc}. After each non-knob metric we list in parentheses which aspect of the experiment are they meant to measure.

\begin{itemize}

\item \oracle \knob Degree of outcome nonlinearity: 0, 1, 2 for linear, non-linear, step-function response surface.

\item \oracle \knob Degree of treatment assignment nonlinearity: 0, 1, 2 for linear, non-linear, step-function treatment assignment model.

\item \oracle \knob Percentage of treated: setting more treated or more control units. Note that despite being an oracle metric, this is in fact easily estimated from data, and is {\em included as a non-oracle measure in the discussion}.

\item \oracle \knob Overlap: binary, whether there was or was not considerable overlap between treated and control.

\item \oracle \knob Alignment: 0,1,2 for no, low, and high alignment between the confounders pertaining to treatment assignment and confounders pertaining to outcome.

\item \oracle \knob Treatment effect heterogeneity: 0, 1, 2 for no, low or high treatment effect heterogeneity. 

\item \oracle Correlation between true propensity score and the outcome (alignment).

\item \oracle Mean Mahalanobis distance to nearest counterfactual neighbor in the ``ground truth'' design matrix (overlap).

\item \oracle Euclidean norm of distance between mean of control and mean of treated in ``ground truth'' design matrix (balance).

\item \oracle Wasserstein distance \citep{villani2008optimal,cuturi2013sinkhorn} between treated and control using the ``ground truth''  design matrix (balance).

\item \oracle $R^2$ of the logit of the true propensity score regressed on the observable design matrix (treatment assignment nonlinearity).

\item \oracle $R^2$ of the true treatment effect regressed on the observable design matrix.

\item \oracle $R^2$ of the outcome regressed on the ``ground truth'' design matrix (outcome nonlinearity).

\item \oracle The ratio of the $R^2$ of the outcome regressed on the observable design matrix divided by the $R^2$ of the outcome regressed on the ``ground truth'' design matrix (outcome nonlinearity).

\item \oracle $R^2$ of the true potential outcome function $Y(0)$ on the control units ($Z=0$) in the observable design matrix (outcome nonlinearity).

\item \oracle $R^2$ of the true potential outcome function $Y(0)$ on the control units ($Z=0$) in the ``ground truth'' design matrix (outcome nonlinearity).

\item \oracle $R^2$ of the true potential outcome function $Y(1)$ on the treated units ($Z=1$) in the observable design matrix (outcome nonlinearity).

\item \oracle $R^2$ of the true potential outcome function $Y(1)$ on the treated units ($Z=1$) in the ``ground truth'' design matrix (outcome nonlinearity).

\item \oracle Standard deviation of the ground truth treatment effect function $\E\left[Y(1)-Y(0)|X\right]$ (treatment effect heterogeneity).

\item $R^2$ of the observed outcome $Y$ on the observed design matrix (outcome nonlinearity).

\item $R^2$ of fitting a propensity score model estimated with logistic regression on the observable design matrix (treatment assignment nonlinearity).

\item $R^2$ between the estimated unit level treatment effect estimated by BART ($\E\left[\hat{Y}(1) - \hat{Y}(0) | X\right]$), and propensity scores estimated with logistic regression on the observable design matrix (alignment).

\item Mean Mahalanobis distance to nearest counterfactual neighbor in the observable design matrix (overlap).

\item Euclidean norm of distance between mean of control and mean of treated in the observable design matrix (balance).

\item Wasserstein distance \citep{villani2008optimal,cuturi2013sinkhorn} between treated and control using the observable design matrix (balance).

\end{itemize}

\subsection{Submissions and Acknowledgements}
\label{sec:appendix:acknowledgements}

We would like to thank the following people for taking the time to submit. Affiliations are those of the first author at the time of submission and methods are in no particular order. Some methods were submitted as being representative in their fields and may not reflect their submitter's beliefs for best practices.

\subsubsection*{Do-It-Yourself Methods}

\begin{center}
  \rowcolors{2}{lightgray}{}
  \begin{tabular}{p{3cm}p{3.5cm}p{6cm}}
    \toprule
    \bfseries Method & Author & Institution \\
    \midrule
    IPTW estimator & Chanmin Kim & Deptartment of Biostatistics, Harvard University \\
    MITSS & Liangyuan Hu and Chenyang Gu & Department of Population Health Science and Policy, Icahn School of Medicine at Mount Sinai \\
    Bayes LM & Christoph Kurz & German Research Center for Environmental Health, Helmholtz Zentrum München \\
    Regression Trees & Dave Harris & Department of Wildlife Ecology and Conservation, University of Florida \\
    Calibrated IPW & Gi-Soo Kim & Department of Statistics, Seoul National University \\
    DR w/GBM + MDIA and Ad Hoc & John Lockwood & Educational Testing Service \\
    Weighted GP & Junfeng Wen and Russ Grenier & Department of Computing Science, University of Alberta \\
    LAS Gen GAM & Leonid Liu and Annie Wang & Analyst Institute \\
    GLM-Boost & Manuel Huber & German Research Center for Environmental Health, Helmholtz Zentrum München \\
    Manual & Mao Hu & Acumen, LLC \\
    RBD TwoStepLM & Peng Ding & Department of Statistics, University of California Berkeley \\
    ProxMatch & Hui Fen Tan, David Miller, and James Savage & Department of Statistics, Cornell University \\
    VarSel NN & Zhipeng Hou and Bryan Keller & Teacher's College, Columbia University \\
    \bottomrule
  \end{tabular}
\end{center}

\subsubsection*{Black Box Methods}

\begin{center}
  \rowcolors{2}{lightgray}{}
  \begin{tabular}{p{3cm}p{3.5cm}p{6cm}}
    \toprule
    \bfseries Method & Author & Institution \\
    \midrule
    MHE Algorithm & Peter Aronow & Department of Political Science, Yale University  \\
    BART & Douglas Galagate and separately Nicole Bohme Carnegie & Department of Math, University of Maryland and Zilber School of Public Health, University of Wisconsin-Milwaukee  \\
    teffects methods & Seth Lirette & Center of Biostatistics and Bioinformatics, University of Mississippi Medical Center  \\
    LASSO+CBPS & James Pustejovsky & Department of Computer Science, Brandeis University  \\
    calCause & Chen Yanover, Omer Weissbrod, Michal Ozery-Flato, Tal El-Hay, Assaf Gottlieb and Yishai Shimoni & IBM Research - Haifa  \\
    BalanceBoost & Qingyuan Zhao & Department of Statistics, Stanford University  \\
    Tree Strat and Adj. Tree Strat & Stefan Wager & Department of Statistics, Stanford University  \\
    h2o Ensemble & Hyunseung Kang & Graduate School of Business, Stanford University  \\
    CBPS & Yongnam Kim & Department of Educational Psychology, University of Wisconsin Madison  \\
    SL+TMLE & Susan Gruber and Mark van der Laan & T.H. Chan School of Public Health, Harvard University  \\
    \bottomrule
  \end{tabular}
\end{center}

\end{document}